\documentclass[12pt,peerreview,draftclsnofoot, letterpaper]{IEEEtran}
\usepackage{amsthm}
\usepackage{amssymb}
\usepackage{graphicx}
\usepackage{multirow}
\usepackage[cmex10]{amsmath}
\usepackage{subfig}
\usepackage{float}
\usepackage{multicol}
\usepackage{balance}
\usepackage{color}

\newcommand{\abs}[1]{\vert #1 \vert }

\newcommand{\defn}{\triangleq}

\newcommand{\mbf}[1]{\mathbf{#1}}

\newcommand{\idc}[1]{\mathfrak{1}\left({#1}\right) }

\newtheorem{lemma}{Lemma}
\newtheorem{cor}{Corollary}
\newtheorem{define}{Definition}

\newtheorem{proposition}{Proposition}
\allowdisplaybreaks
\usepackage{algorithm}
\usepackage{algorithmic}
\IEEEoverridecommandlockouts

\begin{document}
\title{Blind Channel Separation in Massive MIMO System under Pilot Spoofing
 and Jamming Attack}

\author{Ruohan Cao, Tan F. Wong, Hui Gao, Dongqing Wang, and Yueming
  Lu
  \thanks{R.~Cao, H. Gao, D. Wang and Y. Lu are with the Key
    Laboratory of Trustworthy Distributed Computing and Service,
    Ministry of Education, and also the School of Information and
    Communication Engineering, Beijing University of Posts and
    Telecommunications (BUPT), Beijing 100876, China (e-mail:
    \{caoruohan, huigao, wangdongqing, ymlu\}@bupt.edu.cn). }
 \thanks{T. F. Wong is with the Department of Electrical and Computer
    Engineering, University of Florida, FL 32611, U.S.A. (e-mail:
    twong@ufl.edu).}
 \thanks{T. F. Wong was supported by the National Science
    Foundation under Grant CCF-1320086.}
}

\maketitle
\pagestyle{empty}
\begin{abstract}
We consider a channel separation approach to counter the pilot attack
in a massive MIMO system, where malicious users (MUs) perform pilot spoofing and jamming attack (PSJA) in uplink by sending symbols to the basestation (BS) during the channel estimation (CE) phase of the legitimate users (LUs). More specifically, the PSJA strategies employed by the MUs may include (i) sending the random symbols according to arbitrary stationary or non-stationary distributions that are unknown to the BS; (ii) sending the jamming symbols that are correlative to those of the LUs. We analyze the empirical distribution of the received pilot signals (ED-RPS) at the BS, and prove that its characteristic function (CF) asymptotically approaches to the product of the CFs of the desired signal (DS) and the noise, where the DS is the product of the channel matrix and the signal sequences sent by the LUs/MUs. These observations motivate a novel two-step blind channel separation method, wherein we first estimate the CF of DS from the ED-RPS and then extract the alphabet of the DS to separate the channels. Both analysis and simulation results show that the proposed method achieves good channel separation performance in massive MIMO systems.

\end{abstract}

\section{Introduction}
Massive multiple-input multiple-output (MMIMO) systems
\cite{Prasad, Shafi}
exhibit excellent potentials for defense against passive eavesdropping
attacks by using physical-layer security (PLS) techniques~\cite{Liusurvey}.
Many of these PLS techniques however rely on the
knowledge of channel state information (CSI), which is often
estimated in the training phase of a MMIMO system.
More precisely, the amount of CSI available to the system determines the secrecy performance.
Imperfect CSI often courses performance degradation~\cite{chensurvey, Hesurvey}.
 It is thus well known
that MMIMO is vulnerable to active pilot spoofing and jamming attack (PSJA) that aims to
disturb the CSI estimation process~\cite{PLSMaMIMO}. This
PSJA vulnerability presents a weak spot for implementing
PLS techniques in MMIMO.  For instance, in a
time-divison-duplex (TDD) system, a legitimate user (LU) sends pilots to
the base station (BS) during phase of channel estimation (CE). Based on reciprocity between the uplink and
downlink channels, the BS estimates the uplink channel, and uses the
CSI for downlink beamforming. During the CE phase, a malicious user (MU) is free to conduct pilot spoofing attack (PSA) by sending
pilots identical to those of the LU or conduct pilot jamming attack (PJA) by sending any other jamming symbols randomly. This misbehavior
of MU is referred as PSJA.
Due to the existence of PSJA, the BS is beguiled into obtaining a false channel estimate
that is a combination of the legitimate and malicious user channels.
Beamforming with this false CSI in the downlink leaks information to
the MU~\cite{Xiong2015}.

Many signal processing methods have been proposed to counter PSA in MMIMO systems.
In contrast, the work on PJA and PSJA is relatively sparse to the authors' best knowledge.
It is because that the MUs may refuse to
expose their employed pilot sequence set to the BS when performing PJA or PSJA.
The amount of knowledge of attacking has strong impacts on the BS's capability of counteracting attacks.
Some related works are first reviewed as follows.

\subsection{Related works}
\subsubsection{PSA detection}
In refs.~\cite{Kapetanovic2014}--\cite{ICCrandom},
the BS uses PSA detection methods
that
can determine whether a PSA is conducted or
not.
In particular, in
\cite{Kapetanovic2014} and \cite{Kang}, the BS performs PSA
detection by comparing the statistical properties of its observations
with partial CSI known \emph{a priori}.
Refs.~\cite{CFOKnightly} and \cite{CFOZhang} detect PSA according to the existence of carrier frequency
offset (CFO) since CFO naturally exists due to frequency mismatch
between LUs and MUs.
In another family of methods, the LU send random
pilot symbols to the BS~\cite{Tugnait2016}--\cite{Vinogradova}. The random pilot symbols are only known
to the LU, and hence the MUs cannot send the
same pilot symbols. The randomness of the pilot symbols may reduce
the effect of pilot contamination, and allows the BS to detect PSA by determining the number of sources from which its
observations come from.
With sightly difference, ref.~\cite{ICCrandom} places two pilot sequences in one frame that is transmitted to the BS.
The power splitting ratio of the two pilot sequences is known to the BS but not to the MU.
Then, the PSA is detected by comparing the estimation result
obtained by these two sequences.
Refs. \cite{ICTC} and \cite{Xiong2015} give
detection methods performed by LUs. In these methods,
the LUs and the BS respectively estimate the channel using
a two-way training mechanism. The PSA is detected by comparing the
estimation results obtained by the LUs and the BS.

\subsubsection{Channel estimation under PSA}
Refs. \cite{DCETcom}--\cite{TugnaitTcom} focus on estimating the
channels of the LUs and MUs in the presence of PSA.
The
proposed methods all utilize different forms of information or channel \emph{asymmetry}
between the LUs and MUs.
In~\cite{DCETcom} and \cite{Xiong2016}, the channel estimation is
performed by the LUs. If the BS wants to obtain the channel
estimates reliably, it may need to communicate with the LUs via a secure channel that the MUs cannot access.
In~\cite{Bai}--\cite{TugnaitTcomattack}, the BS first uses independent component
analysis (ICA) to separate channels, and then
employ \emph{asymmetry} to match these separated channels with specific
users. To be more specific, in
\cite{Bai} and \cite{RenICC}, it is assumed that some partial CSI (e.g., the path loss
values) is known to the BS \emph{a priori},
and the \emph{a priori} partial CSI of the LU is
different from that of the MU. The BS differentiates the
LU channel from the MU channel by identifying the CSI difference. In~\cite{TugnaitTcom, TugnaitTcomattack}, the \emph{asymmetry} is
based on the restriction that the LU can send
encrypted information to the BS while the MU cannot do so.

\subsubsection{Channel estimation under PJA or PSJA}
In \cite{TTDoPJA},
jamming sequence sent by a single MU is
assumed to {{be}} uniformly distributed over a unit complex-valued set.
Based on this assumption, the BS estimates the channel of the MU
by projecting its observation into the null space of the pilot sequences employed by LUs.
More recently, there are some works on channel estimation under PSJA \cite{CodeOFDM,WangPSJA}, 
where MUs are free to conduct either PSA or PJA.
Both works separate channels of MUs and LUs with well-designed pilot sequence sets,
 and
use different \emph{a priori} partial CSI of MUs and LUs to match these separated channels with
specific users.
In \cite{CodeOFDM}, a pilot sequence is randomly chosen from a code-frequency block group (CFBG) codebook.
As a benefit of CFBG's property, BS firstly determines whether PSA or PJA is conducted.
Then, under PSA, the pilot sequences employed by MUs and LUs can be separated and be used
for channel estimation, but the channel estimation under PJA is not considered in \cite{CodeOFDM}.
In \cite{WangPSJA}, the pilot sequence set contains orthogonal sequences. The BS projects its received superimposed signal into the space
spanned over the orthogonal pilot sequence set. By comparing the projection results in different dimensions,
the pilot sequences of MUs
or LUs can be separated. To facilitate channel estimation,
MUs are assumed to conduct PJA by sending Gaussian noise or conduct PSA
by sending a combination of several pilot sequences with uniform power.

\subsubsection{Summary of related works}

We note that attack detection methods have been investigated extensively \cite{Xiong2015}--\cite{ICTC}.
Some PSA detection methods can be extended for PJA or PSJA detection.
In channel estimation works~\cite{DCETcom}--\cite{WangPSJA},
the MUs are cooperative in the sense that
they send information following certain statistical distributions that are known to the BS and independent with the symbols of LUs \cite{Bai}--\cite{WangPSJA}, or they keep silent when the LUs feed back the estimated results to the BS \cite{DCETcom, Xiong2016}.
As a beneficial result, some efficient methods or criteria, such as ICA or MMSE and etc, can be employed to facilitate channel estimation.
However, in some practical scenarios, the MUs are likely to be incooperative, more powerful and more clever \cite{Trappe}.
More specifically, such MUs are free to send interference symbols according to
arbitrary statistical distribution that is non-stationary and unknown to the BS, and the interference symbols may even depend on
the information of LUs. Also, the MUs may send jamming symbols through all possible communication phases between the LUs and the BS.
For such \emph{powerful} MUs, the performance of existing works may degrade or no longer be provably unbreakable. To this end, new schemes should be proposed.

\subsection{Contributions of this paper}
In this paper, we investigate how to obtain CSI for the BS in the presence of the \emph{powerful} MUs.
Since the powerful MUs may interfere the BS during all possible communication phases between the BS and LUs,
it is important to constitute a channel estimation mechanism that allows the BS to exchange pilot or training information
with the LUs individually. To this end, we focus on separating the channel directions
of the LUs and the MUs.
To be more specific, we consider the following
attack scenarios including the powerful MUs:
\begin{enumerate}
\item  The MUs are free to send the same random symbols as that of LUs, or to send jamming symbols according to arbitrary stationary distributions unknown to the BS, or to vary their used distributions over different transmission instants;
\item  The MUs may overhear the symbols sent by the LUs, and
  send symbols according to their overheard signals;
 \item  The BS does not know which type of attacks is conducted by the MUs.
  \end{enumerate}
Although the separated channel directions are just partial CSI,
with these separated channel directions, the BS is able to receive
information of only one user at a time, and thus eliminate the interference from other non-target users.
Also, the BS is able to
focus its
transmission power towards only one target user at a time.
In summary, the BS can separately receive and transmit to the LUs and MUs
without interference and leakage, which guarantees the performance of
using \emph{asymmetry} configurations (e.g., higher layer authentication protocols, etc) to distinguish between
them, and finally complete full CSI estimation.



{{Notice that part of work is reported in our previous conference paper \cite{Confversion}.
We detail contribution of this paper as follows that make the paper essentially differentiate 
to \cite{Confversion}.
\begin{enumerate}
\item In this paper, we propose a \emph{general} method in the sense that
the proposed method is available to multi-LU against multiple \emph{powerful} malicious
users.
On the contrary,
in  \cite{Confversion}, we only consider a special scenario where
single-MU and single-LU both employ BPSK modulation.
In particular, for the \emph{general} scenario, we propose a blind channel separation method in which the BS
quantizes its observations, and obtains the empirical distribution of
quantized observations for channel separation. It is interesting to note that this
empirical distribution is 
impacted by the channel directions and the distribution of the data
symbols.
As such, the channel directions can be extracted
from the empirical distribution observed by the BS, and hence
achieving blind channel separation.

\item We also analyze the performance of our proposed method in this paper.
On the contrary, performance analysis is not presented in \cite{Confversion}. 
Our analysis work reveals that
as the number of observations and quantization levels approach infinity,
the BS is able to achieve channel separation with nearly errorless performance.
As a beneficial result, our simulation
shows that the directed-to-leakage power ratio achieved by our proposed blind channel
separation method is close to that obtained with perfect CSI.
\end{enumerate}}}
The rest of the paper is organized as follows. The model of the MMIMO
system and the general scenario of spoofing attack will be described
in Section~II. The proposed blind channel separation method will be
detailed in Section~III. Simulation results will be presented in
Section~IV to evaluate the performance of the proposed
method. Finally, conclusions will be drawn in Section~V.

\section{System Model}
\begin{figure}
\centering
\includegraphics[width=0.35\textwidth]{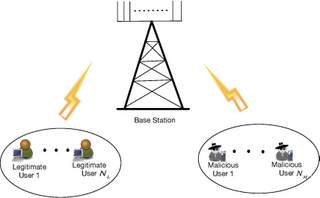}
\caption{System model: the BS is equipped with $M$ antennas. The
  legitimate and malicious users have a single antenna each.}
\label{fig:system_model}
\end{figure}
Consider the system model depicted in Fig.~\ref{fig:system_model}, where a
BS estimates the uplink channels from $N_L$ LUs in the
presence of $N_M$ MUs. The BS is equipped with
$M$
 antennas, while the MUs and
LUs are each equipped with a single antenna. 
For facilitating channel separation in MMIMO system, $M$, $N_L$, and $N_M$
are required to satisfy
$M \gg N_M+N_L$. The uplink and the
corresponding downlink channels satisfy reciprocity.
We perform channel separation to allow beamforming to
individual users over the downlink channels. Assume that the channel
separation process is performed within the coherent time of the
channels. For $j=1,\ldots,N_{L}$ and $k=1,\ldots,N_{M}$, we use the
$M\times 1$ vectors $\boldsymbol{h}_{j}$ and $\boldsymbol{g}_{k}$ to
specify the channels from the $j$th LU and the $k$th
MU to the BS, respectively.
Whenever needed, $\beta_{1, j} $ and
$\beta_{2, k}$ denote the path losses of
channels of the $j$th LU and the $k$th MU,
respectively.
In the uplink, the $j$th
LU and the $k$th MU send random symbols $A_j$
and $B_k$ to the BS,
respectively.
We
allow $A_1, \ldots, A_{N_{L}}$ and $B_1, \ldots, B_{N_{M}}$ to follow
any arbitrary distributions over any finite alphabets.
The distributions of $B_1, \ldots, B_{N_{M}}$ are likely to vary over instants.
Note that since arbitrary symbol distributions are allowed, the proposed scheme will
work for both pilot and data symbols.  We assume that the
distributions of $A_1, \ldots, A_{N_{L}}$ are known to the MUs, while the distributions and even the alphabets of
$B_1, \ldots, B_{N_{M}}$ are, on the contrary, unknown to the BS and
the LUs.

{{For $j=1,\ldots,N_{L}$ and $k=1,\ldots,N_{M}$, let us use $\mathcal{A}_j$ and $\mathcal{B}_k$ 
to denote alphabets of the $j$-th LU and the $k$-th MU, respectively. $\mathsf{a}_j$ and $\mathsf{b}_k$
denote the generic elements of $\mathcal{A}_j$ and $\mathcal{B}_k$, respectively. 
$P_{A_j}\left(\cdot\right)$ and $P_{B_k}\left(\cdot\right)$ are the distributions of $\mathcal{A}_j$ and $\mathcal{B}_k$, respectively. 
$P_{A_{1},\ldots,A_{N_{L}},B_{1},\ldots,B_{N_{M}}}\left(\cdot\right)$ denote joint distribution of $A_{1},\ldots,A_{N_{L}},B_{1},\ldots,B_{N_{M}}$.}}
We assume in each instant, the joint distribution of $A_{1},\ldots,A_{N_{L}}$ and $B_{1},\ldots,B_{N_{M}}$
satisfies $P_{A_{1},\ldots,A_{N_{L}},B_{1},\ldots,B_{N_{M}}}\left(\mathsf{a}_{1},\ldots,\mathsf{a}_{N_{L}},\mathsf{b}_{1},\ldots,\mathsf{b}_{N_{M}}\right)>0$
whenever $P_{A_1}\left(\mathsf{a}_1\right)>0, \ldots P_{A_{N_L}}\left(\mathsf{a}_{N_L}\right)>0$, and  $P_{B_1}\left(\mathsf{b}_1\right)>0, \ldots, P_{B_{N_M}}\left(\mathsf{b}_{N_M}\right)>0$.
This assumption indicates $A_{1},\ldots,A_{N_{L}}$ and $B_{1},\ldots,B_{N_{M}}$ may be dependent with each other,
but arbitrary $B_k$ cannot be exactly determined by other variables.
In practice, it corresponds to the fact that the $k$th MU sends $B_k$
according to its overheard version of $A_{1},\ldots,A_{N_{L}}$. Nevertheless, the $k$th MU cannot get $A_1, \ldots, A_{N_{L}}$ exactly because of the channel fading and noise.


The received symbol of the BS is specified by
{{\begin{equation}
\boldsymbol{y}=\boldsymbol{H}\boldsymbol{A}+\boldsymbol{G}\boldsymbol{B}
+\boldsymbol{w},
 \label{eq:complex_model}
\end{equation}}}
where $\boldsymbol{w}$ is the noise vector, whose elements are
i.i.d. circular-symmetric complex Gaussian (CSCG) random variables
with zero mean and variance $\sigma^{2}$, and
{{$\boldsymbol{H}=\left[\boldsymbol{h}_{1},\boldsymbol{h}_{2},\ldots,\boldsymbol{h}_{N_{L}}\right]$}},
{{$\boldsymbol{G}=\left[\boldsymbol{g}_{1},\boldsymbol{g}_{2},\ldots,\boldsymbol{g}_{N_{M}}\right]$}},
{{$\boldsymbol{A}=\left[A_{1},\ldots,A_{N_{L}}\right]^{T}$}},
and {{$\boldsymbol{B}=\left[B_{1},\ldots,B_{N_{M}}\right]^{T}$}}.

Assuming that the uplink channel described by~\eqref{eq:complex_model} is
used $n$ times within the coherent time of the channel,
for the $i$th instant of use, \eqref{eq:complex_model} gives rise to
{{\begin{equation}\label{eq:ith}
  \boldsymbol{y}_{i}=\boldsymbol{H}\boldsymbol{A}_i+\boldsymbol{G}\boldsymbol{B}_i+\boldsymbol{w}_{i},
\end{equation}}}
for $i=1, 2, \ldots, n$, where $\boldsymbol{A}_i$ is $N_L \times 1$, and
$\boldsymbol{B}_i$ is $N_M\times 1$. 
They are transmitted symbol vectors of the
LUs and the MUs in the $i$th instant, respectively, and
$\boldsymbol{w}_{i}$ is the noisy vector in the $i$th instant.
Stacking the $n$ equations in~\eqref{eq:ith} into a matrix form, we
obtain
{{\begin{equation} \label{eq:matrix_model}
\mathbf{Y}=\left[\boldsymbol{H},\boldsymbol{G}\right]\left[\begin{array}{c}
\mathbf{A}\\
\mathbf{B}
\end{array}\right]+\mathbf{W}
 \end{equation}}}
where {{$\mathbf{Y}
=\left[\boldsymbol{y}_{1},\boldsymbol{y}_{2},\ldots,\boldsymbol{y}_{n}\right]$}},
{{$\mathbf{W}=\left[\boldsymbol{w}_{1},\boldsymbol{w}_{2},\ldots,\boldsymbol{w}_{n}\right]$}},
{{$\mathbf{A}
=\left[\boldsymbol{A}_{1},\boldsymbol{A}_{2},\ldots,\boldsymbol{A}_{n}\right]$}},
and {{$\mathbf{B}=\left[\boldsymbol{B}_{1},\boldsymbol{B}_{2},\ldots,\boldsymbol{B}_{n}\right]$}}.
$\mathbf{A}$ and $\mathbf{B}$ are $N_L \times n$ and $N_M \times n$ matrices, respectively.
In MMIMO systems, it is likely that the columns of
$\left[\boldsymbol{H},\boldsymbol{G}\right]$ are linearly
independent. We make this assumption throughout this paper. Note
that linear independence is the only requirement that we impose on the
channel vectors. Thus correlations among antenna elements are allowed.

The existence of $N_M$ MUs could be detected by attack detection schemes (see \cite{Vinogradova} for example).
Focusing on channel separation, we assume the detection of MUs is perfect.
Reiterating our attack model, $N_M$ and $N_L$ are
known to the BS based on extensively research on attack detection techniques \cite{Xiong2015}--\cite{ICTC}. Following the existing works, we assume that
each user has single antenna and the system works in TDD mode. Unlike the existing techniques, however, the BS does not know the exact distributions of $B_1, \ldots, B_{N_{M}}$,
$\left\{ \beta_{1,1},\ldots\beta_{1,N_{L}}\right\} $,
$\left\{ \beta_{2,1},\ldots\beta_{2,N_{M}}\right\} $,
$\boldsymbol{H}$, or
$\boldsymbol{G}$ \emph{a priori}. In this sense, our proposed channel separation
is blind to the BS.

Let $\boldsymbol{S}$ be the $M\times (N_L+N_M)$ signal subspace
matrix whose columns form an orthonormal basis that spans the column
space of $\left[\boldsymbol{H},\boldsymbol{G}\right]$.
Then, we project $\mathbf{Y}$ onto the signal subspace and get
from~\eqref{eq:matrix_model}
{{\begin{equation}\label{signal_model}
\mathbf{Z}
= \frac{1}{\sqrt{M}}\boldsymbol{S}^{T}\mathbf{Y}
= \left[\mathsf{\boldsymbol{Z}}'_{1},\mathsf{\boldsymbol{Z}}'_{2}\right]\left[\begin{array}{c}
\mathbf{A}\\
\mathbf{B}
\end{array}\right]+\mathbf{N}
\end{equation}}}
where
$\left[{\mathsf{\boldsymbol{Z}}}'_{1},{\mathsf{\boldsymbol{Z}}}'_{2}\right]=\frac{1}{\sqrt{M}}{\boldsymbol{S}}^{T}\left[\boldsymbol{H},\boldsymbol{G}\right]$,
${\mathsf{\boldsymbol{Z}}}'_{1}=\left[{\mathsf{\boldsymbol{z}}}'_{1,1},\ldots,{\mathsf{\boldsymbol{z}}}'_{1,N_{L}}\right]$,
${\mathsf{\boldsymbol{Z}}}'_{2}=\left[{\mathsf{\boldsymbol{z}}}'_{2,1},\ldots,{\mathsf{\boldsymbol{z}}}'_{2,N_{M}}\right]$,
${{\mathbf{N}}}=\frac{1}{\sqrt{M}}{\boldsymbol{S}}^{T}\mathbf{W}$,
and the elements of ${{\mathbf{N}}}$ are i.i.d. Gaussian random
variables with zero mean and variance $\frac{\sigma^{2}}{M}$.
Clearly, ${{\mathbf{N}}}$ is independent of
$\left[\mathsf{\boldsymbol{Z}}'_{1},
  \mathsf{\boldsymbol{Z}}'_{2}\right] \left[\begin{array}{c}
                                              \mathbf{A}\\
                                              \mathbf{B}
\end{array}\right]$.
It is argued in~\cite{Muller2014} that this independence and the
Gaussianity of ${{\mathbf{N}}}$ imply that
\begin{equation}\label{estimation_channel}
  \left[\boldsymbol{\hat H},\boldsymbol{\hat G}\right]
  = \sqrt{M}\boldsymbol{S}\left[\mathsf{\boldsymbol{Z}}'_{1},\mathsf{\boldsymbol{Z}}'_{2}\right]
\end{equation}
would be a reasonable estimator for the channel pair
$\left[\boldsymbol{H},\boldsymbol{G}\right]$ if
$\left[\mathsf{\boldsymbol{Z}}'_{1},\mathsf{\boldsymbol{Z}}'_{2}\right]$
could be found.  In practice $\boldsymbol{S}$ is not known \emph{a
  priori}, but can be estimated from the singular value decomposition
$\mathbf{Y}=\boldsymbol{U}\boldsymbol{\varSigma}\boldsymbol{V}^{T}$
with orthogonal matrices $\boldsymbol{U}\in\mathbb{R}^{M\times M}$,
$\boldsymbol{V}\in\mathbb{R}^{n\times n}$ and $M\times n$ diagonal
singular value matrix $\boldsymbol{\varSigma}$.  Then $\boldsymbol{S}$
can be well approximated by the first $(N_L+N_M)$ left singular vectors
in $\boldsymbol{U}$ when $M$ is large in the MMIMO
system~\cite{Muller2014}.

{{Incidentally, we can show that
$\left[\mathsf{\boldsymbol{Z}}'_{1},\mathsf{\boldsymbol{Z}}'_{2}\right]$
cannot be uniquely determined from $\mathbf{Y}$ without additional authentication, or the knowledge of
$\left\{ \beta_{1,1},\ldots\beta_{1,N_{L}}\right\} $,
$\left\{ \beta_{2,1},\ldots\beta_{2,N_{M}}\right\} $,
$\mathbf{A}$, and $\mathbf{B}$.
Notice that when $\mathbf{A}$ and $\mathbf{B}$ have the same
distribution in PSA, swapping the columns of
$\left[ \mathsf{\boldsymbol{Z}}'_{1},
  \mathsf{\boldsymbol{Z}}'_{2}\right]$ by observing $\mathbf{Z}$
in~\eqref{signal_model} would not change the distribution of
$\mathbf{Z}$. Thus, it is impossible to obtain any decision rule among the permutations of the columns of
$\left[ \mathsf{\boldsymbol{Z}}'_{1},
  \mathsf{\boldsymbol{Z}}'_{2}\right]$.}}

As will be argued in the next section, it is however possible to
separate channels in the sense of estimating
{{$\left[\frac{\boldsymbol{h}_{1}}{\abs{\boldsymbol{h}_{1}}},\ldots,\frac{\boldsymbol{h}_{N_{L}}}{\abs{\boldsymbol{h}_{N_{L}}}},\frac{\boldsymbol{g}_{1}}{\abs{\boldsymbol{g}_{1}}},\ldots,\frac{\boldsymbol{g}_{N_{M}}}{\abs{\boldsymbol{g}_{N_{M}}}}\right]$}}
via~\eqref{estimation_channel}, up to a permutation of the columns and a phase ambiguity on each column. 
Interchangeably, we also term the above-mentioned separation as channel direction separation, because the obtained $\left[\frac{\boldsymbol{h}_{1}}{\abs{\boldsymbol{h}_{1}}},\ldots,\frac{\boldsymbol{h}_{N_{L}}}{\abs{\boldsymbol{h}_{N_{L}}}},\frac{\boldsymbol{g}_{1}}{\abs{\boldsymbol{g}_{1}}},\ldots,\frac{\boldsymbol{g}_{N_{M}}}{\abs{\boldsymbol{g}_{N_{M}}}}\right]$ characterizes the channel directions of all users.
As a result, we will be able
to separate the downlink beamforming directions from the BS to the LUs
and MUs based on channel reciprocity.

\section{Blind Channel Separation} \label{sec:bcs}
First, notice that the columns of
{{$\left[\mathsf{\boldsymbol{Z}}'_{1},\mathsf{\boldsymbol{Z}}'_{2}\right]
\left[\begin{array}{c}
        \mathbf{A}\\
        \mathbf{B}
\end{array}\right]$}}
are $(N_L+N_M)\times 1$ random vectors
that range over the alphabet
{{$ \mathcal{Z}=\left\{
  \sum_{j=1}^{N_{L}}\mathsf{a}_{j}\mathsf{\boldsymbol{z}}'_{1,j} +
  \sum_{k=1}^{N_{M}}\mathsf{b}_{k}\mathsf{\boldsymbol{z}}'_{2,k}:
  \mathsf{a}_{j}\in\mathcal{A}_{j},\mathsf{b}_{k}\in\mathcal{B}_{k}\right\}$}},
where $\mathcal{A}_{j}$ and $\mathcal{B}_{k}$ are the respective
alphabets of ${A}_{j}$ and ${B}_{k}$, for $j=1,\ldots,N_{L}$ and
$k=1,\ldots,N_{M}$. The main idea of our channel separation scheme is
to use $\mathcal{Z}$ to obtain
$\left[\mathsf{\boldsymbol{Z}}'_{1},\mathsf{\boldsymbol{Z}}'_{2}\right]$
up to a column permutation, and then achieve the desired separation of
channel directions via~\eqref{estimation_channel}. In this section,
we will first illustrate how to separate channels from perfect $ \mathcal{Z}$,
then propose estimation of $ \mathcal{Z}$ based on received observations, finally 
give a method indicating channel separations from received observations step-by-step.

\subsection{Channel separation from perfect knowledge of $ \mathcal{Z}$}
For easy description of 
our proposed method, we give two definitions.
{{\begin{define}
For $\boldsymbol{v}\in\mathcal{Z}$,
$\boldsymbol{v}'\in\mathcal{Z}$, $\boldsymbol{v}\neq\boldsymbol{v}'$,
if one vector $\mathsf{\boldsymbol{z}}'\notin\mathcal{Z}$ satisfies
\begin{equation}\label{def:cover}
\frac{\left(\boldsymbol{v}-\boldsymbol{v}'\right)^{H}}{\left|\boldsymbol{v}-\boldsymbol{v}'\right|\left|\mathsf{\boldsymbol{z}}'\right|}\mathsf{\boldsymbol{z}}'=\exp\left(\mathsf{i}\theta\right),
\end{equation}where $\theta$ could be arbitrary angle,
we refer to this property as $\mathsf{\boldsymbol{z}}'$ covers $\boldsymbol{v}$ and $\boldsymbol{v}'$.
\end{define}
\begin{define}
For $\mathsf{\boldsymbol{z}}'\neq0$, $\mathsf{\boldsymbol{z}}'\notin\mathcal{Z}$, subset $\mathcal{Z}_{s}\subset\mathcal{Z}$, 
if arbitrary $\boldsymbol{v}\in\mathcal{Z}_{s}$, $\forall \boldsymbol{v}'\in\mathcal{Z}_{s}$ such that
$\mathsf{\boldsymbol{z}}'$ covers $\boldsymbol{v}$ and $\boldsymbol{v}'$, 
we refer to this property as $\mathsf{\boldsymbol{z}}'$ covers $\mathcal{Z}_{s}$ or $\mathsf{\boldsymbol{z}}'$ covers $\abs{\mathcal{Z}_{s}}$
points of $\mathcal{Z}$.
\end{define}
According to Definition 2, if $\mathsf{\boldsymbol{z}}'$ covers $\mathcal{Z}_{s}$ and $\mathcal{Z}_{s}=\mathcal{Z}$, 
we term that as $\mathsf{\boldsymbol{z}}'$ covers $\mathcal{Z}$.}}

Then, to explain how we may obtain
{{$\left[\mathsf{\boldsymbol{Z}}'_{1},\mathsf{\boldsymbol{Z}}'_{2}\right]$}}
from $\mathcal{Z}$, let us start by considering an example with
$N_L=N_M=1$ and both the LU and MU send BPSK
symbols. That is,
{{${\mathsf{\boldsymbol{Z}}}'_{1}=\left[{\mathsf{\boldsymbol{z}}}'_{1,1}\right]$}},
{{${\mathsf{\boldsymbol{Z}}}'_{2}=\left[{\mathsf{\boldsymbol{z}}}'_{2,1}\right]$}},
and
{{$\mathcal{Z} = \left\{
  \underbrace{\mathsf{\boldsymbol{z}}'_{1,1}+\mathsf{\boldsymbol{z}}'_{2,1}}_{{\boldsymbol{v}}_{A}} ,
  \underbrace{-\mathsf{\boldsymbol{z}}'_{1,1}+\mathsf{\boldsymbol{z}}'_{2,1}}_{{\boldsymbol{v}}_{B}} ,
  \underbrace{\mathsf{\boldsymbol{z}}'_{1,1}-\mathsf{\boldsymbol{z}}'_{2,1}}_{{\boldsymbol{v}}_{C}} ,
  \underbrace{-\mathsf{\boldsymbol{z}}'_{1,1}-\mathsf{\boldsymbol{z}}'_{2,1}}_{{\boldsymbol{v}}_{D}}
  \right\}$}}.
\begin{figure}
\centering
\includegraphics[width=0.35\textwidth]{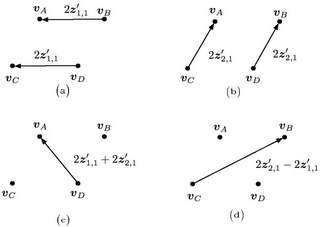}
\caption{Subfigures (a) and (b) show that all points of
  $\mathcal{Z}$ lie on lines along the directions of
  $\mathsf{\boldsymbol{z}}'_{1,1}$ and
  $\mathsf{\boldsymbol{z}}'_{2,1}$, respectively. Subfigures (c) and
  (d) show that only two points in  $\mathcal{Z}$ lie on lines along the
  directions of
  $\mathsf{\boldsymbol{z}}'_{1,1}+\mathsf{\boldsymbol{z}}'_{2,1}$ and
  $\mathsf{\boldsymbol{z}}'_{2,1}-\mathsf{\boldsymbol{z}}'_{1,1}$,
  respectively.}
\label{fig:binaryalphabet}
\end{figure}
Then, it is clear that
{{\begin{align}
\label{connect1}
{\boldsymbol{v}}_{A}-{\boldsymbol{v}}_{B} &=2 \mathsf{\boldsymbol{z}}'_{1,1},
\\
\label{connect2}
 {\boldsymbol{v}}_{C}-{\boldsymbol{v}}_{D} &=2 \mathsf{\boldsymbol{z}}'_{1,1}.
\end{align}}}
According to Definition 2, (\ref{connect1}) and (\ref{connect2}) jointly indicate that $\mathsf{\boldsymbol{z}}'$ covers $\mathcal{Z}$.
Similarly,  it is easy to see that ${\boldsymbol{v}}_{A}-{\boldsymbol{v}}_{C} =2 \mathsf{\boldsymbol{z}}'_{2,1}$,
$ {\boldsymbol{v}}_{B}-{\boldsymbol{v}}_{D} =2
                                             \mathsf{\boldsymbol{z}}'_{2,1}$,
and hence $\mathsf{\boldsymbol{z}}'_{2,1}$ also covers $\mathcal{Z}$.
On the other hand, {{as long as $\mathsf{\boldsymbol{z}}'_{1,1}$ and $\mathsf{\boldsymbol{z}}'_{2,1}$ are linearly independent}}, we have that only
{{${\boldsymbol{v}}_{A}-{\boldsymbol{v}}_{D} =
2\mathsf{\boldsymbol{z}}'_{1,1} + 2\mathsf{\boldsymbol{z}}'_{2,1}$}}
and
{{${\boldsymbol{v}}_{B}-{\boldsymbol{v}}_{C}=2\mathsf{\boldsymbol{z}}'_{2,1}-2\mathsf{\boldsymbol{z}}'_{1,1}$}}.
Thus, neither
$\mathsf{\boldsymbol{z}}'_{1,1} + \mathsf{\boldsymbol{z}}'_{2,1}$ nor
$\mathsf{\boldsymbol{z}}'_{2,1}-\mathsf{\boldsymbol{z}}'_{1,1}$ covers $\mathcal{Z}$.  Note that the above cases exhaust the differences
between all pairs of points in $\mathcal{Z}$. In summary, among all
the pairwise differences, each of $\{\mathsf{\boldsymbol{z}}'_{1, 1},
\mathsf{\boldsymbol{z}}'_{2,1}\}$ covers $\mathcal{Z}$, but
each of $\{\mathsf{\boldsymbol{z}}'_{1,1} + \mathsf{\boldsymbol{z}}'_{2,1},
\mathsf{\boldsymbol{z}}'_{2,1}-\mathsf{\boldsymbol{z}}'_{1,1}\}$ does not. As a result, we may obtain $\mathsf{\boldsymbol{z}}'_{1, 1}$ and
$\mathsf{\boldsymbol{z}}'_{2,1}$ from $\mathcal{Z}$ by finding out all
pairwise differences that cover $\mathcal{Z}$. It turns out that this
observation extends to the general case as summarized in the
proposition below:
\begin{proposition} \label{prop1}
Consider the general alphabet
{{$
 \mathcal{Z} = \left\{
  \sum_{j=1}^{N_{L}}\mathsf{a}_{j}\mathsf{\boldsymbol{z}}'_{1,j} +
  \sum_{k=1}^{N_{M}}\mathsf{b}_{k}\mathsf{\boldsymbol{z}}'_{2,k}
  :\mathsf{a}_{j}\in\mathcal{A}_{j},\mathsf{b}_{k}\in\mathcal{B}_{k}
  \right\}
$}}
where $\mathcal{A}_{j}$ and $\mathcal{B}_{k}$, for $j=1, \ldots, N_L$
and $k=1, \ldots, N_M$, are finite and with cardinalities at least
$2$. {{As long as columns of $\{ \mathsf{\boldsymbol{z}}'_{1,1},
  \ldots,\mathsf{\boldsymbol{z}}'_{1,N_{L}},
  \mathsf{\boldsymbol{z}}'_{2,1}, \ldots,
  \mathsf{\boldsymbol{z}}'_{2,N_{M}} \}$ are linearly independent,}}
every vector in
$\left\{ \mathsf{\boldsymbol{z}}'_{1,1},
  \ldots,\mathsf{\boldsymbol{z}}'_{1,N_{L}},
  \mathsf{\boldsymbol{z}}'_{2,1}, \ldots,
  \mathsf{\boldsymbol{z}}'_{2,N_{M}} \right\}$ covers $\mathcal{Z}$.
On the contrary, each vector of the form
$\sum_{j=1}^{N_{L}}c_{1,j}\mathsf{\boldsymbol{z}}'_{1,j} +
\sum_{k=1}^{N_{M}}c_{2,k}\mathsf{\boldsymbol{z}}'_{2,k}$, where at
least two coefficients in
$\left\{ c_{1,1}, \ldots, c_{1,N_{L}}, c_{2,1}, \ldots,
  c_{2,N_{M}}\right\}$ are nonzero, does not cover $\mathcal{Z}$.

\end{proposition}
\begin{IEEEproof}
See the Appendix A.
\end{IEEEproof}
{{Notice that in massive MIMO system, columns of $\left[\boldsymbol{H},\boldsymbol{G}\right]$ are 
linearly independent in probability. It indicates that columns of $\{ \mathsf{\boldsymbol{z}}'_{1,1},
  \ldots,\mathsf{\boldsymbol{z}}'_{1,N_{L}},
  \mathsf{\boldsymbol{z}}'_{2,1}, \ldots,
  \mathsf{\boldsymbol{z}}'_{2,N_{M}} \}$ are almost always linearly independent.}}

Proposition~\ref{prop1} allows us to obtain the columns of
$[\mathsf{\boldsymbol{Z}}'_{1}, \mathsf{\boldsymbol{Z}}'_{2}]$ by
finding all the pairwise differences of vectors in $\mathcal{Z}$ that
covers $\mathcal{Z}$ itself. 
The steps given below show how Proposition 1 works on ${\mathcal{Z}}$. We use P1, P2 and P3 to label steps in this perfect case.
\begin{enumerate}
\item[P1] Let us write ${\mathcal{Z}}=\left\{ \boldsymbol{v}_{1},\ldots,\boldsymbol{v}_{\left|\mathcal{Z}\right|}\right\}$,
obtain the set of pairwise differences
  $\mathcal{D} =\big\{
    {\boldsymbol{v}}_{i}-{\boldsymbol{v}}_{j} : i< j
    \text{ and } i, j \in \{1,\ldots,|{\mathcal{Z}}|
    \}\big\}$.
\item[P2] {{Find subsets 
   $\mathcal{D}^{*}$, $\mathcal{D}_{1}$, $\ldots$, $\mathcal{D}_{T}$, 
   these subsets simultaneously satisfy following conditions.
 {{  \begin{align*}
  & \mathcal{D}^* = \left\{ \boldsymbol{d}_{1}, \boldsymbol{d}_{2},
    \ldots, \boldsymbol{d}_{T}\right\} \subseteq\mathcal{D}\\
  & \mathcal{D}_{t} = \left\{ \boldsymbol{d}\in\mathcal{D} :
    \left|\frac{\boldsymbol{d}^{H}\boldsymbol{d}_{t}}{\big |
          \boldsymbol{d}\big | \big |
          \boldsymbol{d}_{t}\big | }-1\right|=0 \right\}, t=1,2\ldots, T\\
          &\mathcal{D}=\bigcup_{t=1}^{T}\mathcal{D}_{t}, \mathcal{D}_{s} \cap \mathcal{D}_{t} = \emptyset, s \neq t.
   \end{align*}}}
Notice that  $\mathcal{D}_t$ depends on $\boldsymbol{d}_{t}$. It is the $t$th element of $\mathcal{D}^{*}$.}}     
For each $t \in \{1, 2, \ldots, T\}$, if
 ${\boldsymbol{v}}_{i} -
 {\boldsymbol{v}}_{j}\in\mathcal{D}_{t}$, then we collect
 ${\boldsymbol{v}}_{i}$ and ${\boldsymbol{v}}_{j}$ in
${\mathcal{Z}}_{t}$.
\item[P3]  Define the weight of
  $\boldsymbol{d}_{t}$ as
  $W\left(\boldsymbol{d}_{t}\right)=\left|{\mathcal{Z}}_{t}\right|$.
  Use the $N_L+N_M$ vectors in $\mathcal{D}^*$ with the largest
  weights as estimates of the columns of
$[\mathsf{\boldsymbol{Z}}'_{1}, \mathsf{\boldsymbol{Z}}'_{2}]$.
\end{enumerate}
{{In step P1, all pairwise differences of $\mathcal{Z}$ are obtained. Then,
in step P2, we get ${\mathcal{Z}}_{1}, \ldots, {\mathcal{Z}}_{T}$. They are covered by 
pairwise differences according to Definition 2. Finally, step P3 separates channels
by finding $N_L+N_M$ vectors, each of which covers the most points.
It exploits Proposition 1 
that only pairwise differences along directions of channels could cover most points.}}

The above-mentioned steps are implemented for the perfect case that we have ${\mathcal{Z}}$.
Let us go back to practical where ${\mathcal{Z}}$ is unknown. To separate channels, we need to
to estimate ${\mathcal{Z}}$ from the observation $\mathbf{Y}$ given
in~\eqref{eq:matrix_model}, which will be discussed in next subsection.


\subsection{Estimation of $\mathcal{Z}$ based on $\mathbf{Y}$}
To estimate $\mathcal{Z}$, we first consider the case that the columns of
$\left[\mathsf{\boldsymbol{Z}}'_{1},\mathsf{\boldsymbol{Z}}'_{2}\right]
\left[\begin{array}{c}
        \mathbf{A}\\
        \mathbf{B}
\end{array}\right]$
are i.i.d. vectors in order to introduce our method of estimating
$\mathcal{Z}$. In Proposition~\ref{prop2}, we will prove that the
proposed method is also applicable to the case of non-i.i.d. columns.
Further notice that the columns of $\mathbf{N}$
in~\eqref{signal_model} are i.i.d. random vectors that have the same
distribution as that of the generic $(N_M+N_L)\times 1$ random vector
$\mathbf{n}$, whose elements are independent Gaussian random variables
with zero mean and variance $\frac{\sigma^{2}}{M}$.  If the columns
of
$\left[\mathsf{\boldsymbol{Z}}'_{1},\mathsf{\boldsymbol{Z}}'_{2}\right]
\left[\begin{array}{c}
        \mathbf{A}\\
        \mathbf{B}
\end{array}\right]$
are i.i.d. random vectors that have the same distribution as the
generic $(N_M+N_L)\times 1$ random vector $\mathbf{{z}}'$, then
the columns of $\mathbf{Z}$, given in~\eqref{signal_model}, are
i.i.d. random vectors that have the same distribution as that of
$\mathbf{z}=\mathbf{z}'+\mathbf{n}$.  Let $F_{\mathbf{z}}$,
$F_{\mathbf{z}'}$, and $F_{\mathbf{n}}$ denote the distributions of
${\mathbf{z}}$, ${\mathbf{z}'}$, and ${\mathbf{n}}$,
respectively. Then, because $\mathbf{z}'$ and $\mathbf{n}$ are
independent, we have
\begin{equation} \label{eq:chi} \Phi_{F_{\mathbf{z}}}
  (\boldsymbol{\omega}) = \Phi_{F_{\mathbf{z}'}}(\boldsymbol{\omega})
  \cdot \Phi_{F_{\mathbf{n}}}(\boldsymbol{\omega}).
\end{equation}
where $\Phi_F( \boldsymbol{\omega})$ denotes the characteristic
function of the distribution $F$, and
$\boldsymbol{\omega} = [\omega_1, \ldots, \omega_{2N_L+2N_M}]^T$ is the $(2N_L+2N_M)\times 1$
frequency vector. Note that the noise variance parameter $\sigma^2$ is
a characteristic of the receiver circuitry and can be measured
\emph{a priori}. We may assume that its value is known, and thus
$\Phi_{F_{\mathbf{n}}}(\boldsymbol{\omega})
  =\exp\left\{ -\frac{\sigma^{2}}{4M}\left| \boldsymbol{\omega}\right|^{2}\right\}
$
is also known. On the other hand, $F_{\mathbf{z}}$ can be approximated
by the empirical distribution of $\mathbf{Z}$ obtained directly from
the observation $\mathbf{Y}$ as in~\eqref{signal_model}. Hence the
distribution $F_{\mathbf{z}'}$ of $\mathbf{z}'$ can be estimated
using~\eqref{eq:chi}.

For ease of discussion, let us also use $\mathbf{z}$ to denote a
generic column in the matrix $\mathbf{Z}$.  To estimate
$F_{\mathbf{z}'}$ efficiently, we quantize
${\mathbf{z}} = [z_1, \ldots, z_{N_L+N_M}]^T$ and use FFT to obtain the characteristic function of the quantized
version of ${\mathbf{z}}$ as follows.
Consider $m_1$ quantization levels
$\mathsf{\widetilde{u}}_{1},\mathsf{\widetilde{u}}_{2},\ldots,\mathsf{\widetilde{u}}_{m_1}$,
and the corresponding quantization intervals
$\mathcal{B}\left(\mathsf{\widetilde{u}}_{1}\right),
\mathcal{B}\left(\mathsf{\widetilde{u}}_{2}\right), \ldots,
\mathcal{B}\left(\mathsf{\widetilde{u}}_{m_1}\right)$:
{{\begin{align}
&-\alpha_1=\mathsf{\widetilde{u}}_{1}<\mathsf{\widetilde{u}}_{2}<\mathsf{\widetilde{u}}_{3}<
  \cdots <\mathsf{\widetilde{u}}_{m_1-1} \leq \alpha_1 = \mathsf{\widetilde{u}}_{m_1},
\nonumber \\
&\mathcal{B}\left(\widetilde{\mathsf{u}}_{j}\right)=\begin{cases}
\begin{array}{cc}
\left(-\infty,\,\widetilde{\mathsf{u}}_{1}\right], & j=1\\
\left(\widetilde{\mathsf{u}}_{j-1},\,\widetilde{\mathsf{u}}_{j}\right],
  & j=2,3,\ldots,m_1-1\\
\left(\widetilde{\mathsf{u}}_{m_1-1},\,+\infty\right), & j=m_1
\end{array}\end{cases}
\label{q_u1}
\end{align}}}
where $\alpha_1 > 0$,
$\widetilde{\mathsf{u}}_{j}-\widetilde{\mathsf{u}}_{j-1}=\delta_1$
for $j=2,3,\ldots,m_1$.  Hence, we have
$m_1 = \left\lceil \frac{2 \alpha_1}{\delta_1} \right\rceil + 1$.
The elements $z_1, \ldots, z_{N_L+N_M}$ are respectively quantized to
$\widetilde z_1, \ldots, \widetilde z_{N_L+N_M}$ according to:
{{\begin{align} \label{eq:quant}
\widetilde{z}_{i}=
\sum_{j=1}^{m_{1}}\widetilde{\mathsf{u}}_{j}\,\idc{\Re\left(z_{i}\right)\in\mathcal{B}\left(\widetilde{\mathsf{u}}_{j}\right)}+
\mathsf{i}\sum_{j=1}^{m_{1}}\widetilde{\mathsf{u}}_{j}\,\idc{\Im\left(z_{i}\right)\in\mathcal{B}\left(\widetilde{\mathsf{u}}_{j}\right)},
 i=1,\ldots, N_L+N_M,
\end{align}}}where $\idc{\cdot}$ denotes the indicator function.  The quantized
version of $\mathbf{z}$ is then
$\widetilde{\mathbf{z}}=\left[\widetilde z_1, \ldots, \widetilde z_{N_L+N_M}\right]^T$.
Write
$\widetilde{\mathcal{U}} =
\left\{\mathsf{\widetilde{u}}_{1},\mathsf{\widetilde{u}}_{2},\ldots,\mathsf{\widetilde{u}}_{m_1}
\right\}$, then the alphabet of $\widetilde{\mathbf{z}}$ is
$\widetilde{\mathcal{U}}^{(N_{L}+N_{M})}\times\mathsf{i}\widetilde{\mathcal{U}}^{(N_{L}+N_{M})}$.  We will denote it and enumerate its
elements as
$\widetilde{\mathcal{Z}} = \left\{
  \widetilde{\mathsf{\boldsymbol{u}}}_{1},
  \widetilde{\mathsf{\boldsymbol{u}}}_{2}, \ldots,
  \mathsf{\widetilde{\boldsymbol{u}}}_{{m^{2N_L+2N_M}_1}}\right\}$.

Let
$\widetilde{\mathbf{Z}} = \left[ \widetilde{\mathbf{z}}_1,
  \widetilde{\mathbf{z}}_2, \ldots, \widetilde{\mathbf{z}}_n \right]$,
where the $i$th column $\widetilde{\mathbf{z}}_i$ is the quantized
version of the $i$th column of $\mathbf{Z}$. Next, obtain the
empirical probability mass function (pmf) of the columns of
$\widetilde{\mathbf{Z}}$ as
\begin{equation}\label{f2}
  \triangle {F}_{\widetilde{\ensuremath{\mathbf{z}}}}
  \left(\widetilde{\mathsf{\boldsymbol{u}}}_{j}\right)
  =
  \frac{1}{n} \sum_{i=1}^{n} \idc{\widetilde{\ensuremath{\mathbf{z}}}_{i}=\widetilde{\mathsf{\boldsymbol{u}}}_{j}}
\end{equation}
for {{$j=1,2,\ldots, m^{2N_L+2N_M}_1$}}.  Thus, the characteristic
function of the empirical distribution
$\triangle {F}_{\widetilde{\ensuremath{\mathbf{z}}}}$ is
{{\begin{align}\label{eq:FT}
\Phi_{\triangle {F}_{\widetilde{\ensuremath{\mathbf{z}}}}}\left( \boldsymbol{\omega}\right)
=\sum_{j=1}^{m_{1}^{2N_{L}+2N_{M}}}\triangle{F}_{\widetilde{\ensuremath{\mathbf{z}}}}\left(\widetilde{\mathsf{\boldsymbol{u}}}_{j}\right)\exp\left\{ -\mathsf{i}\left[\Re\left\{ \widetilde{\mathsf{\boldsymbol{u}}}_{j}^{T}\right\} ,\Im\left\{ \widetilde{\mathsf{\boldsymbol{u}}}_{j}^{T}\right\} \right]\boldsymbol{\omega}\right\}
\end{align}}}
where $\mathsf{i} = \sqrt{-1}$.

Similarly, let
$\widetilde{\mathbf{Z}}' = \left[ \widetilde{\mathbf{z}}'_1,
  \widetilde{\mathbf{z}}'_2, \ldots, \widetilde{\mathbf{z}}'_n
\right]$, where the $i$th column $\widetilde{\mathbf{z}}'_i$ is the
quantized version of the $i$th column of
{{$\left[ \mathsf{\boldsymbol{Z}}'_{1},
     \mathsf{\boldsymbol{Z}}'_{2}\right]
\left[\begin{array}{c}
\mathbf{A}\\
\mathbf{B}
\end{array}\right]$}}.
To do this quantization step, we employ an extended quantization
alphabet
{{$\widetilde{\mathcal{Z}}' = \left\{
  \widetilde{\mathsf{\boldsymbol{v}}}_{1},
  \widetilde{\mathsf{\boldsymbol{v}}}_{2}, \ldots,
  \mathsf{\widetilde{\boldsymbol{v}}}_{{m^{(2N_L+2N_M)}_2}}\right\}$}},
which is obtained by employing the same uniform quantization
in~\eqref{q_u1} with $m_2$ quantization levels covering a range
$\left(-\alpha_{2},\alpha_{2}\right)$. The length of each quantization
duration is $\delta_2$. Hence, we have $m_2 = \left\lceil \frac{2 \alpha_2}{\delta_2} \right\rceil + 1$.
To facilitate analysis, we assume $\widetilde{\mathcal{Z}}'$ is an extension of
$\widetilde{\mathcal{Z}}$, i.e.,
$\widetilde{\mathcal{Z}} \subset \widetilde{\mathcal{Z}}'$.  Then, the
empirical pmf of the columns of $\widetilde{\mathbf{Z}}'$ is
{{\begin{equation}\label{fz'}
  \triangle {F}_{\widetilde{\ensuremath{\mathbf{z}}}'}
  \left(\widetilde{\mathsf{\boldsymbol{v}}}_{j}\right)
  =
  \frac{1}{n} \sum_{i=1}^{n} \idc{\widetilde{\ensuremath{\mathbf{z}}}'_{i}=\widetilde{\mathsf{\boldsymbol{v}}}_{j}}
\end{equation}}}
for $j=1,2,\ldots, m^{2N_L+2N_M}_2$.

Using
$\Phi_{\triangle {F}_{\widetilde{\ensuremath{\mathbf{z}}}}}\left(
  \boldsymbol{\omega}\right)$ in place of
$\Phi_{F_{\mathbf{z}}} \left( \boldsymbol{\omega}\right)$ and
$\triangle {F}_{\widetilde{\ensuremath{\mathbf{z}}}'}$ in place of
${F}_{\mathbf{z}'}$ in~\eqref{eq:chi}, we obtain the estimator
\begin{equation}\label{eq:hat}
  \triangle \widehat{F}_{\widetilde{\ensuremath{\mathbf{z}}}'}
  = \Phi^{-1} \left(
    \frac{\Phi_{\triangle {F}_{\widetilde{\ensuremath{\mathbf{z}}}}}\left( \boldsymbol{\omega}\right)}{
      \Phi_{F_{\mathbf{n}}} \left(\boldsymbol{\omega}\right)} \right)
\end{equation}
for $\triangle {F}_{\widetilde{\ensuremath{\mathbf{z}}}'}$, where
$\Phi^{-1}$ denotes the inverse Fourier Transform with respect
to~\eqref{eq:FT}. Note that the forward and inverse Fourier Transforms
in~\eqref{eq:hat} can be efficiently implemented using FFT and inverse
FFT (IFFT), respectively. The asymptotic accuracy of this estimator is
addressed by
Proposition~\ref{prop2} below:
\begin{proposition} \label{prop2}
  By
  choosing $n = \Omega \left(m_2^{2(N_L+N_M)} \right)$,
  $m_2 = \Omega \left(m_1^3\right)$,
 $\alpha_{1}=\alpha_{2}$, $2\alpha_{1}=(m_1-1)\delta_1$, $2\alpha_{2}=(m_2-1)\delta_2$, and $\frac{m_{2}-1}{m_{1}-1}
 $ is  integer, there exists
  $\epsilon_{m_2} \rightarrow 0$ as
  $\alpha_1 \rightarrow \infty$,
  $m_2 \rightarrow \infty$, $\delta_1 \rightarrow 0$, and $\delta_2 \rightarrow 0$ such that
whether $\mathbf{Z}$ is i.i.d. or non-i.i.d., as long as $\Re\left\{ \mathcal{Z}\right\}\subseteq\left[-\alpha_{1},\alpha_{1}-\delta_2\right]^{N_{M}+N_{L}}$,
$\Im\left\{ \mathcal{Z}\right\} \subseteq\left[-\alpha_{1},\alpha_{1}-\delta_2\right]^{N_{M}+N_{L}}$, then the proposed estimator  (\ref{eq:hat}) has
{{ $ \left|\triangle {F}_{\widetilde{\ensuremath{\mathbf{z}}}'}\left(
      \widetilde{\boldsymbol{v}}_{j}\right)
    -\triangle \widehat{F}_{\widetilde{\ensuremath{\mathbf{z}}}'}\left(
      \widetilde{\boldsymbol{v}}_{j}\right)\right|
  \leq \epsilon_{m_2}$}}
for $j=1,2,\ldots, m^{N_L+N_M}_2$.
\end{proposition}
\begin{IEEEproof}
See the Appendix B.
\end{IEEEproof}$\left[-\alpha_{1},\alpha_{1}-\delta_2\right]^{N_{M}+N_{L}}$ indicates
Cartesian product of $\left[-\alpha_{1},\alpha_{1}-\delta_2\right]$ with ${N_{M}+N_{L}}$ times.
The assumptions of $\Re\left\{ \mathcal{Z}\right\}\subseteq\left[-\alpha_{1},\alpha_{1}-\delta_2\right]^{N_{M}+N_{L}}$
and $\Im\left\{ \mathcal{Z}\right\} \subseteq\left[-\alpha_{1},\alpha_{1}-\delta_2\right]^{N_{M}+N_{L}}$
indicate that $\mathcal{Z}$ must be included in the quantization range. This could be achieved
as $\alpha_1 \rightarrow \infty$, while the transmitted power of MUs and LUs are not infinite. 

Now, notice that $\mathcal{Z}$ is the support set of the empirical
distribution, $\triangle {F}_{\mathbf{z}'}$, of the columns of
$\left[ \mathsf{\boldsymbol{Z}}'_{1},
  \mathsf{\boldsymbol{Z}}'_{2}\right] \left[\begin{array}{c}
                                              \mathbf{A}\\
                                              \mathbf{B}
\end{array}\right]$.
Clearly, as the quantization is fine enough,
$\triangle {F}_{\widetilde{\ensuremath{\mathbf{z}}}'}$ approximates
$\triangle {F}_{\mathbf{z}'}$ well. Thus, Proposition~\ref{prop2} tells us
that we may use the essential support set
\[
  \widehat{\mathcal{Z}} = \left\{
  \widetilde{\boldsymbol{u}}\in\widetilde{\mathcal{Z}}:
  \triangle\widehat{F}_{\widetilde{\ensuremath{\mathbf{z}}}'}\left(
    \widetilde{\boldsymbol{u}}\right) > \epsilon_{m_2}
\right\}
\]
of $\triangle \widehat{F}_{\widetilde{\ensuremath{\mathbf{z}}}'}$ to
estimate $\mathcal{Z}$.


It is worth noting that Proposition~\ref{prop2} does not require the columns of
$\widetilde{\mathbf{Z}}'$ to be i.i.d. random vectors as detailed in the proof. Thus, Proposition~\ref{prop2} extends the attack model to
allow the MUs' symbols to be arbitrary distributed, as
discussed before.

\emph{Remark 1.}
Proposition 2 empowers our proposed method
to be applicable even in the presence of the powerful MUs,
who send
symbol sequences following arbitrary unknown distribution.
On contrast, some existing channel separation or estimation
works based on maximum likelihood theory are no longer
useful in the presence of the powerful MUs, since
these works need to know the distributions of MUs' symbols
\emph{a priori}.

\emph{Remark 2.}
In addition, we
use only the support of
$\triangle \widehat{F}_{\widetilde{\ensuremath{\mathbf{z}}}'}$ to
estimate $\mathcal{Z}$.
Correlations between the columns of $\mathbf{Z}$ are
immaterial to the proposed scheme. It indicates that our proposed
scheme allows the symbols of MUs to statistically depend on
the symbols of LUs.  This property differs our work
with ICA-based methods, which require symbols of MUs and LUs to be independent with each other.


\subsection{Blind channel separation method} \label{sec:algo}
According to Propositions~\ref{prop1} and~\ref{prop2},
practical channel separation {{method}} could be achieved.
In particular, Proposition~\ref{prop2} is used for estimating ${\mathcal{Z}}$.
In other words, $ \widehat{\mathcal{Z}}$ is obtained.
Then, Proposition~\ref{prop1} achieves channel separation based on $ \widehat{\mathcal{Z}}$.

\begin{figure}
\centering
\includegraphics[width=0.35\textwidth]{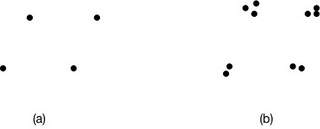}
\caption{An example of comparison between ${\mathcal{Z}}$ and $\widehat{\mathcal{Z}}$. They are illustrated in subfigures (a) and (b), respectively. }
\label{fig:alphbets}
\end{figure}

Combining Propositions~\ref{prop1} and~\ref{prop2}, we can obtain the
following practical channel separation method:
\begin{enumerate}
\item Perform SVD on the observation matrix $\mathbf{Y}$, collect
  the first $N_L+N_M$ left singular vectors as columns of
  $\mathbf{S}$, and obtain $\mathbf{Z} = \frac{1}{\sqrt{M}}
  \mathbf{S}^T \mathbf{Y}$.
\item Quantize the columns of $\mathbf{Z}$ to obtain
  $\widetilde{\mathbf{Z}} = \left[ \widetilde{\mathbf{z}}_1,
    \widetilde{\mathbf{z}}_2, \ldots, \widetilde{\mathbf{z}}_n
  \right]$, based on~\eqref{eq:quant}. Obtain the empirical pmf
  $\triangle {F}_{\widetilde{\ensuremath{\mathbf{z}}}}$ of the columns
  of $\widetilde{\mathbf{Z}}$ as described in~\eqref{f2}.
\item Employ~\eqref{eq:hat} to obtain
  $\triangle \widehat{F}_{\widetilde{\ensuremath{\mathbf{z}}}'}$ via
  FFT and IFFT.
\item Choose a small $\epsilon > 0$. Obtain the essential support set
$ \widehat{\mathcal{Z}} = \left\{
  \widetilde{\boldsymbol{u}}\in\widetilde{\mathcal{Z}}:
  \triangle\widehat{F}_{\widetilde{\ensuremath{\mathbf{z}}}'}\left(
    \widetilde{\boldsymbol{u}}\right) > \epsilon
\right\}$.
\item  Write
  $\widehat{\mathcal{Z}} = \left\{ \widehat{\boldsymbol{v}}_{1},
    \widehat{\boldsymbol{v}}_{2}, \ldots,
    \widehat{\boldsymbol{v}}_{\left|\widehat{\mathcal{Z}}\right|}\right\}$.
Obtain the set of pairwise differences
  $\mathcal{D} =\left\{
    \widehat{\boldsymbol{v}}_{i}-\widehat{\boldsymbol{v}}_{j} : i< j
 \text{ and } i, j \in \{1,\ldots,|\widehat{\mathcal{Z}}|
    \}\right\}$.
   \item Choose a small $\gamma > 0$. {{Find subsets 
   $\mathcal{D}^{*}$, $\mathcal{D}_{1}$, $\ldots$, $\mathcal{D}_{T}$, 
   these subsets simultaneously satisfy following conditions.
  {{ \begin{align*}
  & \mathcal{D}^* = \left\{ \boldsymbol{d}_{1}, \boldsymbol{d}_{2},
    \ldots, \boldsymbol{d}_{T}\right\} \subseteq\mathcal{D}\\
  & \mathcal{D}_{t} = \left\{ \boldsymbol{d}\in\mathcal{D} :
    \left|\frac{\boldsymbol{d}^{H}\boldsymbol{d}_{t}}{\big |
          \boldsymbol{d}\big | \big |
          \boldsymbol{d}_{t}\big | }-1\right|\leq\gamma \right\}, t=1,2\ldots, T\\
          &\mathcal{D}=\bigcup_{t=1}^{T}\mathcal{D}_{t}, \mathcal{D}_{s} \cap \mathcal{D}_{t} = \emptyset, s \neq t.
   \end{align*}}}
Notice that  $\mathcal{D}_t$ depends on $\boldsymbol{d}_{t}$. It is the $t$th element of $\mathcal{D}^{*}$.
Algorithm 1 is gave for illustrating how to find such $\mathcal{D}_1, \mathcal{D}_2, \ldots,  \mathcal{D}_T$, and $\mathcal{D}^{*}$.}}   
   For each $t \in \{1, 2, \ldots, T\}$, if
  $\widehat{\boldsymbol{v}}_{i} -
  \widehat{\boldsymbol{v}}_{j}\in\mathcal{D}_{t}$, then we collect
  $\widehat{\boldsymbol{v}}_{i}$ and $\widehat{\boldsymbol{v}}_{j}$ in
  $\widehat{\mathcal{Z}}_{t}$. Algorithm 1 illustrates how to implement this step.
\item  Define the weight of
  $\boldsymbol{d}_{t}$ as
  $W\left(\boldsymbol{d}_{t}\right) =
  \sum_{\widehat{\boldsymbol{v}}\in\widehat{\mathcal{Z}}_{t}}\triangle
  \widehat{F}_{\widetilde{\ensuremath{\mathbf{z}}}'}
  \left(\widehat{\boldsymbol{v}} \right)$.
  Use the $N_L+N_M$ vectors in $\mathcal{D}^*$ with the largest
  weights as estimates of the columns of
$[\mathsf{\boldsymbol{Z}}'_{1}, \mathsf{\boldsymbol{Z}}'_{2}]$.
\item Employ~\eqref{estimation_channel} to estimate
{{$ \left[\frac{\boldsymbol{h}_{1}}{\abs{\boldsymbol{h}_{1}}},
    \ldots,
    \frac{\boldsymbol{h}_{N_{L}}}{\abs{\boldsymbol{h}_{N_{L}}}},
    \frac{\boldsymbol{g}_{1}}{\abs{\boldsymbol{g}_{1}}}, \ldots,
    \frac{\boldsymbol{g}_{N_{M}}}{\abs{\boldsymbol{g}_{N_{M}}}}\right]$}}
  up to a permutation of the columns and up to a phase ambiguity on
  each column, from the estimated
 {{$[\mathsf{\boldsymbol{Z}}'_{1}, \mathsf{\boldsymbol{Z}}'_{2}]$}}
  obtained in 7).
\end{enumerate}
Steps 1)-4) are designed to obtain the estimator in (\ref{eq:hat}), and
the estimation of ${\mathcal{Z}}$. The performance of steps 1)-4) is
guaranteed by Proposition 2.
Steps 5)-7) are obtained by respectively modifying steps P1-P3 a little bit.
Compared to steps P1-P3, in step 5), $\mathcal{D}$ is obtained from $\widehat{\mathcal{Z}}$ rather than ${\mathcal{Z}}$.
Note that we use $\widehat{\mathcal{Z}}$ to approximate ${\mathcal{Z}}$ for practical application.
$\widehat{\mathcal{Z}}$ often contains many more points than
$\mathcal{Z}$ as shown in Fig.~\ref{fig:alphbets}.  To solve this problem, step 6) first clusters all
pairwise differences by defining {{$\mathcal{D}_{t} = \left\{ \boldsymbol{d}\in\mathcal{D} :
    \left|\frac{\boldsymbol{d}^{H}\boldsymbol{d}_{t}}{\big |
          \boldsymbol{d}\big | \big |
          \boldsymbol{d}_{t}\big | }-1\right|\leq\gamma \right\}$}}. Step~ 7) uses a likelihood metric to
approximately obtain the set of covering pairwise difference vectors
among the clusters obtained in step~6).
{\small{\begin{algorithm}[!h]  
    \caption{Achieving $\widehat{\mathcal{Z}}_{1}, \widehat{\mathcal{Z}}_{2}, \ldots, \widehat{\mathcal{Z}}_{T}$ by step 6)} 
    \begin{algorithmic}[1]
        \STATE $\mathcal{D}=\left\{ \boldsymbol{d}'_{1},\boldsymbol{d}'_{2}\ldots\boldsymbol{d}'_{\left|\mathcal{D}\right|}\right\}$,  $\widehat{\mathcal{Z}} = \left\{ \widehat{\boldsymbol{v}}_{1},
    \widehat{\boldsymbol{v}}_{2}, \ldots,
    \widehat{\boldsymbol{v}}_{\left|\widehat{\mathcal{Z}}\right|}\right\}$, $t=1$.
    \STATE $\boldsymbol{d}_{1}= \boldsymbol{d}'_{1}$, $\mathcal{D}^* = \left\{ \boldsymbol{d}_{1}\right\}$.
        \WHILE{$\mathcal{D}\neq\emptyset$}
       \STATE  $\mathcal{D}_t=\emptyset$, $\widehat{\mathcal{Z}} _t=\emptyset$
        \FOR{$j=1$ to $\left|\mathcal{D}\right|$}  
        \IF{$ \left|\frac{\boldsymbol{d}'^{H}_{j}\boldsymbol{d}_{t}}{\big |
          \boldsymbol{d}'^{H}_{j}\big | \big |
          \boldsymbol{d}_{t}\big | }-1\right|\leq\gamma$}  
        \STATE $\mathcal{D}_{t}\impliedby\boldsymbol{d}'_{j}$   
        \ENDIF 
        \STATE $j\leftarrow j+1$  
         \ENDFOR
          \FOR{$i=1$ to $\left|\widehat{\mathcal{Z}}\right|$}  
             \FOR{$k=i+1$ to $\left|\widehat{\mathcal{Z}}\right|$} 
        \IF{ $\widehat{\boldsymbol{v}}_{i} -
  \widehat{\boldsymbol{v}}_{k}\in\mathcal{D}_{t}$}  
        \STATE $\widehat{\mathcal{Z}}_{t}\impliedby\widehat{\boldsymbol{v}}_{i}$, $\widehat{\mathcal{Z}}_{t}\impliedby\widehat{\boldsymbol{v}}_{k}$ 
        \ENDIF 
        \STATE $k\leftarrow k+1$         
         \ENDFOR
          \STATE $i\leftarrow i+1$ 
         \ENDFOR
         \STATE $\mathcal{D}\leftarrow\mathcal{D}-\mathcal{D}_t$, $\mathcal{D}=\left\{ \boldsymbol{d}'_{1},\boldsymbol{d}'_{2}\ldots\boldsymbol{d}'_{\left|\mathcal{D}\right|}\right\}$,$t\leftarrow t+1$, $\boldsymbol{d}_{t}= \boldsymbol{d}'_{1}$, $\mathcal{D}^* \impliedby \left\{ \boldsymbol{d}'_{1}\right\}$.
         \ENDWHILE
 \end{algorithmic}  
\end{algorithm}}}

\emph{Remark 3} The proposed method obtains
partial CSI. Comparing with full CSI, i.e., $\left[\boldsymbol{H},\boldsymbol{G}\right]$, the obtained result still has
the permutation of columns and the phase ambiguity in
  each column. This observation is similar to some blind source separation methods (e.g., ICA), and needs
  further processing to obtain full CSI. Nevertheless,
 unlike the existing methods, it is clear that the proposed method
does not require the \emph{a priori} knowledge of $\mathcal{Z}$ or the symbol
alphabets of the LUs and MUs, and thus imposes no restriction
on the statistic dependence between symbols of the LUs and MUs. 


\subsection{Complexity analysis}
We proceed to briefly analyze the computational complexity of the
blind channel separation method described in Section~\ref{sec:algo}
as follows:
\begin{enumerate}
\item The SVD operations, which has a complexity
  $\mathcal{O}\left(\max\{ M^{2}n, n^3\}\right)$, dominates in
  step~1).
\item Quantizing the columns of $\boldsymbol{Z}$ and obtaining
  $\triangle {F}_{\widetilde{\ensuremath{\mathbf{z}}}}$ require
  $\mathcal{O}\left(nMm_{1}\right)$ complexity in step~2).
\item The FFT (and IFFT) {{operations}} in step~3) perform
  $(N_L+N_M)$-dimensional $m_2$-point FFT. Hence the complexity of
  step~3) is
  $\mathcal{O}\left(\left(N_{L}+N_{M}\right)m_{2}^{\left(N_{L}+N_{M}\right)}\log
    m_{2}\right)$.
\item[4)-8)] Notice that
  $\left|\mathcal{Z}\right|=\prod_{i=1}^{N_{L}}\left|\mathcal{A}_{i}\right|\prod_{j=1}^{N_{M}}\left|\mathcal{B}_{j}\right|$.
  By Proposition~\ref{prop2},
  $\left|\widehat{\mathcal{Z}}\right| \approx
  \left|\mathcal{Z}\right|$ for large $m_2$ and $n$. In addition,
  $T \leq \left|\mathcal{D}\right| \leq
  \left|\widehat{\mathcal{Z}}\right|^2 \approx
  \left|\mathcal{Z}\right|^2$. Hence, the complexity of steps 4)--7)
  can be approximately upper bounded by
  $\mathcal{O}\left( \left(N_{L}+N_{M}\right)
    \left(\prod_{i=1}^{N_{L}}
      \left|\mathcal{A}_{i}\right|\prod_{j=1}^{N_{M}}\left|\mathcal{B}_{j}\right|\right)^{3}
  \right)$.
\end{enumerate}
Recall that {{$m_2 = 
\Omega(m^3_1)$}} and we would often choose a large $m_2$
that gives
$\left(\prod_{i=1}^{N_{L}} \left|\mathcal{A}_{i}\right|
  \prod_{j=1}^{N_{M}} \left|\mathcal{B}_{j}\right|\right)^{3}\leq
m_{2}^{\left(N_{L}+N_{M}\right)}\log m_{2}$. Therefore, the total
complexity of the blind channel separation method can be
characterized by 
{{\begin{equation*}
 \mathcal{O}\left(\max\{ M^{2}n, n^3\}\right)
  + \mathcal{O}\left(nMm_{2}\right) + \mathcal{O}\left(\left(N_{L}+N_{M}\right)m_{2}^{\left(N_{L}+N_{M}\right)}\log m_{2}\right).
\end{equation*}}}
Notice that the computation complexity of the proposed method is
higher than the existing methods. For example, the complexity
of FastICA is no more than $\mathcal{O}\left(\max\{M^{2}n,n^{3}\}\right)+\mathcal{O}\left(nM^{2}m_{2}\right)$\cite{FastICAJ, FastICAC}.
 Nevertheless, ICA methods require the \emph{a priori} knowledge of $\mathcal{Z}$ and statistic independence
 between symbols of the LUs and MUs, while the proposed method does not require that.
In other words, the proposed method imposes less {{restrictions}} on MUs at cost of its computation complexity.
In that sense, the proposed method {{may be}} more suitable to combat against the MUs
whose misbehavior is {{out of the control of the BS}},  {{as long as the computation capability of the BS is sufficient}}. 

\section{Performance Evaluation}

In this section, we present {{our}} simulation results to evaluate the
performance of the blind channel separation (BCS) method described in
Section~\ref{sec:algo}.   In the simulation, we
generate $100,000$ instances of the channel vectors based on the block
Rayleigh fading model. That is, the elements of each column in
$\boldsymbol{H}$ and $\boldsymbol{G}$ are chosen as i.i.d. CSCG random variables with mean
$0$ mean and variance $1$. We then average performance metric, given later, over $100,000$ channel {{realizations}}.  We choose
$\alpha_1=\alpha_2$ and  $m_1=m_2=64$.  We consider the antenna array sizes of  $M=64$ and
$M=128$, and two different numbers of observation {{instances}}, namely,
$n=300$ and $n=500$, respectively. 
Each LU sends symbols with power of $P_s$.
The ratio $\frac{P_s}{\sigma^2}$ is
the per-antenna signal-to-noise ratio (SNR) of the LU's
signal. We vary the SNR value in the simulation to evaluate the
channel separation performance of the proposed method.

{{\subsection{Performance metric}
Let $\boldsymbol{h}$ be a column of
$\left[ \boldsymbol{H},\boldsymbol{G}\right]$, and
$\mbf{P}_{\boldsymbol{h}}=\frac{\boldsymbol{h}\boldsymbol{h}^{H}}{\left|\boldsymbol{h}\right|^{2}}$
be the projection on the subspace spanned
by $\boldsymbol{h}$.  Consider the $j$th legitimate user. Suppose
that $\boldsymbol{d}$ is an uplink channel direction vector from this user
to the BS obtained by some channel estimation algorithm. By
reciprocity, downlink beamforming is performed based on
$\boldsymbol{d}$.  Then,
$\abs{\mbf{P}_{\boldsymbol{h}_j} \boldsymbol{d}}^2$ is the power
directed to the $j$th legitimate user by the BS. On the other
hand, the total power leaked to other legitimate users and malicious
users is given by
$\sum_{l=1, l \neq j}^{N_L} \abs{\mbf{P}_{\boldsymbol{h}_{l}}
  \boldsymbol{d}}^2 + \sum_{k=1}^{N_M}
\abs{\mbf{P}_{\boldsymbol{g}_{k}} \boldsymbol{d}}^2$.  Hence, the
directed-to-leakage power ratio (DLPR)
\begin{equation} \label{eq:DLPRatio}
  \frac{\abs{\mbf{P}_{\boldsymbol{h}_j}
      \boldsymbol{d}}^2}{\sum_{l=1, l \neq j}^{N_L}
    \abs{\mbf{P}_{\boldsymbol{h}_{l}} \boldsymbol{d}}^2 +
    \sum_{k=1}^{N_M} \abs{\mbf{P}_{\boldsymbol{g}_{k}}
      \boldsymbol{d}}^2}
\end{equation}
measures the ratio of the power directed to the $j$th legitimate
user to the power leaked to all others if $\boldsymbol{d}$ is
employed to perform beamforming based on channel reciprocity.
Clearly, substituting $\boldsymbol{d} =\frac{\boldsymbol{h}_{j}}{\left|\boldsymbol{h}_{j}\right|}$
in~\eqref{eq:DLPRatio} obtains the DLPR value when the channel direction
estimation is perfect.

Now, let
$\widehat{\boldsymbol{H}} = \left[ \widehat{\boldsymbol{h}}_1,
  \ldots, \widehat{\boldsymbol{h}}_{N_L+N_M} \right]$ be the channel
direction vectors estimated using the blind channel separation scheme
described in Section~\ref{sec:algo}. Since we do not know which column
of $\widehat{\boldsymbol{H}}$ corresponds to $\boldsymbol{h}_j$, we
consider the maximum among the DLPRs of all the possibilities:
\begin{align*} 
 \hspace*{-5pt}
\text{DLPR}_j(\boldsymbol{H}, \boldsymbol{G})\defn 
\max_{i \in \{1,2, \ldots, N_L+N_M\}} 
\frac{\abs{\mbf{P}_{\boldsymbol{h}_j}
      \widehat{\boldsymbol{h}}_i}^2}{\sum_{l=1, l \neq j}^{N_L}
    \abs{\mbf{P}_{\boldsymbol{h}_{l}} \widehat{\boldsymbol{h}}_i}^2 +
    \sum_{k=1}^{N_M} \abs{\mbf{P}_{\boldsymbol{g}_{k}}
      \widehat{\boldsymbol{h}}_i}^2}.
\end{align*}
Finally, we conservatively use the worst-case DLPR among all the
legitimate users as our performance metric:
\begin{equation} \label{eq:DLPR}
 \text{DLPR}(\boldsymbol{H}, \boldsymbol{G}) \defn
 \min_{j \in \{1,2, \ldots, N_L\}} \text{DLPR}_j(\boldsymbol{H},
 \boldsymbol{G}).
\end{equation}
Note that the DLPR in~\eqref{eq:DLPR} is a function of the channel
vectors $[\boldsymbol{H}, \boldsymbol{G}]$.}}

{
{{\subsection{Simulation}
We first consider a PSA scenario with two LUs and one MU.
The MU conducts PSA by sending equally likely random BPSK symbols as those of LUs.
The BPSK symbols of all these three users are independent with each other.
We simulate DLPR achieved by perfect CSI, our proposed BCS scheme, and traditional ICA \cite{FastICAJ}, 
respectively. The length of observations is set to $n=300$. Two array sizes of $M=128$ and $M=64$
are considered. Fig.~\ref{fig:PSA} shows the obtained result. 
Notice that it is a standard scenario for ICA since all users send independent symbols.
We observe from Fig.~\ref{fig:PSA} that our proposed BCS outperforms ICA, and achieves 
near-optimal performance. The performance improvement is coursed by the fact that our proposed BCS
attempts to cut down the impact of noise. In particular,
revisiting \eqref{eq:chi}, our proposed BCS estimates distribution of desired signals (i.e., $\left[\mathsf{\boldsymbol{Z}}'_{1},\mathsf{\boldsymbol{Z}}'_{2}\right]
\left[\begin{array}{c}
        \mathbf{A}\\
        \mathbf{B} 
\end{array}\right]$) by removing CF of noise from CF of received noisy observations (i.e., $\boldsymbol{Y}$). 
On the other hand,  ICA is derived for desired signals without any noise, and then applied to the noisy observations
straightforwardly.

Then, we consider a PJA scenario with two LUs and one MU.
The LUs send equally likely random BPSK symbols, while the MU
conducts PJA by sending PAM symbols according to its wiretapped signal from
the LUs. In particular, 
$P_{B_1|A_1}(+3|+1)=P_{B_1|A_1}(-1|+1)=\frac{1}{3}$, $P_{B_1|A_1}(+1|+1)=P_{B_1|A_1}(-3|+1)=\frac{1}{6}$, 
$P_{B_1|A_1}(+1|-1)=P_{B_1|A_1}(+3|-1)=\frac{1}{6}$, $P_{B_1|A_1}(-1|-1)=P_{B_1|A_1}(-3|-1)=\frac{1}{3}$, where $P_{B_1|A_1}$
denotes the conditional probability of symbol of the MU given symbol sent by the first LU. The length of observations is set to $n=800$.
We observe from the Fig.~\ref{fig:PJA1} that for $M=64$ and $M=128$, the proposed method achieves DLPR close to
that achieved by perfect CSI
within $1$dB. 
In contrast,
even the per-antenna SNR reaches $16$dB, the DLPR achieved by the traditional ICA scheme is only about 50\%
of that achieved by the scheme with perfect CSI. 
This performance gap is brought by the fact that ICA requires independence between symbols 
sent by users. In this scenario, the dependence between symbols of these three users
degrades the performance. 
Our proposed BCS works based on alphabet estimation, imposing no restrict on 
statistic dependence between symbols of users. Therefore, our proposed BCS outperforms
ICA.

Finally, we simulate another PJA scenario with two LUs and two MUs. 
The MUs employ non-stationary PAM attack. In particular, in odd instants, 
they send random symbols following statistic distribution $P_{B_1}(+3)=P_{B_2}(+3)=\frac{1}{3}$,  $P_{B_1}(-1)=P_{B_2}(-1)=\frac{2}{3}$.
In even instants, they send random symbols following statistic distribution $P_{B_1}(+1)=P_{B_2}(+1)=\frac{2}{3}$,  $P_{B_1}(-1)=P_{B_2}(-1)=\frac{1}{3}$.
As shown in Fig.~\ref{fig:PJA2}, for $M=64$ and $M=128$, our proposed BCS still outperforms traditional ICA.
ICA separates channels by minimizing a contrast function that measure gaussianity of its output. 
Its employed contrast function may be not always suitable for the non-stationary attack. As a result,
the performance is degraded. On the contrast, Proposition 2 guarantees our proposed BCS
is still available to non-stationary attack.}}


 \begin{figure}[t]
\centering
\includegraphics[width=0.40\textwidth]{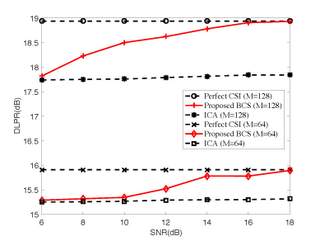}
\caption{DLPR performance under PSA scenario with $N_L = 2$, $N_M = 1$,
 and all users send BPSK symbols.}
\label{fig:PSA}
\end{figure}

 \begin{figure}[t]
\centering
\includegraphics[width=0.40\textwidth]{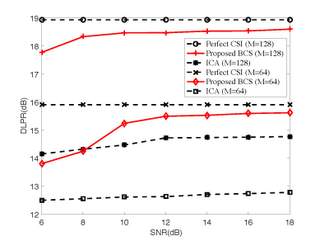}
\caption{DLPR performance under PJA scenario where the malicious user sends 
symbols correlated to those of legitimate users.}
\label{fig:PJA1}
\end{figure}

 \begin{figure}[t]
\centering
\includegraphics[width=0.40\textwidth]{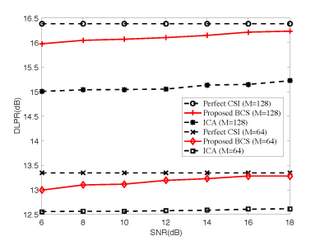}
\caption{DLPR performance under PJA scenario where the malicious user sends symbols
according to varying distribution.}
\label{fig:PJA2}
\end{figure}

All the above results demonstrate that the proposed BCS method is effective in separating and
estimating the channel directions of the LU and the MU.
With our {proposed} method, if the MUs conduct attack,
the attack signals will expose the MUs' channel directions.
Then, forced by our proposed method, the MUs may have to keep silent, such that the BS cannot estimate
their channel vectors. Nevertheless, in such case, without interference of the MUs, the BS
can get the CSI of the LUs.
As a result, the BS is able to focus its power along the direction of the LUs,
while the leakage to the MUs is little due to quasi-orthogonality of the channel vectors in MMIMO system.
In summary, the MUs cannot achieve its goal of eavesdropping the downlink information.
As a beneficial result, the security of the system is guaranteed.


\section{Conclusions}
We have proposed a BCS method to differentiate
the channel directions from the LU and MU to the BS
in the uplink of a MMIMO system.  With channel reciprocity, the
BS is able to use the channel directions to steer the beam toward the
LUs and MUs separately with minimal power and
information leakage, and further verify their identities by use of
some higher layer authentication protocols. Simulation results have shown
that the proposed method can achieve good channel separation
performance in terms of achieving good DLPR performance at low to
moderate per-antenna SNR and near-perfect DLPR performance at
high SNR. It is also noted that even when the MUs are allowed to
impersonate the LUs by sending symbols of distribution
identical to that of the LUs' symbols, the proposed scheme still works very well. Moreover, the proposed
method does not requires the BS to have any partial CSI of the
channels \emph{a priori}. The complexity of the method is
polynomial in the number of antennas and the number of observation
instants, but is exponential in the number of users. Thus the
method is most suitable for the practical use when the number of users
is small. As an extension to the current work, it is interesting to
investigate wheatear the knowledge of the LUs' symbols (such as
pilot symbols) can be utilized to reduce the complexity of the channel
separation method.

\appendices{}
\section{Proof of Proposition 1}
\begin{IEEEproof}
  Let
  $\mathcal{A}_{j}=\left\{
    \mathsf{a}_{j,1},\ldots,\mathsf{a}_{j,\left|\mathcal{A}_{j}\right|}\right\}$
  and
  $\mathcal{B}_{k}=\left\{
    \mathsf{b}_{k,1},\ldots,\mathsf{b}_{k,\left|\mathcal{B}_{k}\right|}\right\}$.
   Without loss of generalization, we focus on the
  direction of $\mathsf{\boldsymbol{z}}'_{1,j}$. For an arbitrary
   point  $\boldsymbol{u}$ in
$\mathcal{Z}$, we may write it as
$ \boldsymbol{u} =  \mathsf{a}_{j,s}\mathsf{\boldsymbol{z}}'_{1,j} +
 \mathsf{\boldsymbol{z}}'$,
for some $s \in \{1,\ldots,\left|\mathcal{A}_{j}\right|\}$, and
$\mathsf{\boldsymbol{z}}'=\sum_{j'=1,j\neq j'}^{N_{L}}\mathsf{a}_{j'}\mathsf{\boldsymbol{z}}'_{1,j'}+\sum_{k=1}^{N_{M}}\mathsf{b}_{k}\mathsf{\boldsymbol{z}}'_{2,k}$.
Then there must exist another point $\boldsymbol{u}' \in \mathcal{Z}$
satisfying
$\boldsymbol{u}' =
\mathsf{a}_{j,s'}\mathsf{\boldsymbol{z}}'_{1,1}+\mathsf{\boldsymbol{z}}'$
with some $s'\neq s$. Because the difference
\[
  \boldsymbol{u}' -\boldsymbol{u} =
(\mathsf{a}_{j,s'}-\mathsf{a}_{j,s})\mathsf{\boldsymbol{z}}'_{1,j},
\]
$\boldsymbol{u}$ and $\boldsymbol{u}'$ are covered by
$\mathsf{\boldsymbol{z}}'_{1,j}$ according to Definition 1.
It is obvious that this same
argument applies to every point in $\mathcal{Z}$ and every direction
in
$\left\{ \mathsf{\boldsymbol{z}}'_{1,1},
  \ldots,\mathsf{\boldsymbol{z}}'_{1,N_{L}},
  \mathsf{\boldsymbol{z}}'_{2,1}, \ldots,
  \mathsf{\boldsymbol{z}}'_{2,N_{M}} \right\}$. Hence, every vector in
$\left\{ \mathsf{\boldsymbol{z}}'_{1,1}, \ldots,
  \mathsf{\boldsymbol{z}}'_{1,N_{L}}, \mathsf{\boldsymbol{z}}'_{2,1},
  \ldots, \mathsf{\boldsymbol{z}}'_{2,N_{M}}\right\}$ covers
$\mathcal{Z}$.

Furthermore, without loss of generalization, consider the vector
$c_{1,j}\mathsf{\boldsymbol{z}}'_{1,j} +
c_{2,k}\mathsf{\boldsymbol{z}}'_{2,k}$.
Focus on the point
\[
  \boldsymbol{v} = \mathsf{a}_{j,\hat{s}}\mathsf{\boldsymbol{z}}'_{1,j} +
\mathsf{b}_{k,\hat{t}}
\mathsf{\boldsymbol{z}}'_{2,k} + \bar{\mathsf{\boldsymbol{z}}}' \in
\mathcal{Z}
\]
where
$\bar{\mathsf{\boldsymbol{z}}}'=\sum_{j'=1,j\neq j}^{N_{L}}\mathsf{a}_{j'}\mathsf{\boldsymbol{z}}'_{1,j'}+\sum_{k'=1,k'\neq k}^{N_{M}}\mathsf{b}_{k'}\mathsf{\boldsymbol{z}}'_{2,k'}$. $\mathsf{a}_{j,\hat{s}}$ and $\mathsf{b}_{k,\hat{t}}$ are chosen according to
$\hat{s}=\underset{s\in\mathcal{S}}{\arg}\min\Im\left\{ \frac{c_{2}}{c_{1}}\mathsf{a}_{j,s}\right\}$,
where {{$\mathcal{S}=\left\{ s:\arg\min_{s\in\left\{ 1,2,\ldots,\left|\mathcal{A}_{j}\right|\right\} }\Re\left\{ \frac{c_{2}}{c_{1}}\mathsf{a}_{j,s}\right\} \right\} $}}
and
\[
\hat{t}=\underset{t\in\mathcal{T}}{\arg}\max\Im\left\{ \mathsf{b}_{k,t}\right\}.
\]where {$\mathcal{T}=\left\{ t:{\arg\max}_{t\in\left\{ 1,2,\ldots,\left|\mathcal{B}_{k}\right|\right\}}\Re\left\{ \mathsf{b}_{k,t}\right\} \right\} $}.

If there were a different
$\boldsymbol{v}' \in \mathcal{Z}$ such that $\boldsymbol{v}'$ and
$\boldsymbol{v}$ are covered by $c_{1,j}\mathsf{\boldsymbol{z}}'_{1,j} +
c_{2,k}\mathsf{\boldsymbol{z}}'_{2,k}$, {{notice that columns of {\small{$\{ \mathsf{\boldsymbol{z}}'_{1,1},
  \ldots,\mathsf{\boldsymbol{z}}'_{1,N_{L}}
  \mathsf{\boldsymbol{z}}'_{2,1}, \ldots,
  \mathsf{\boldsymbol{z}}'_{2,N_{M}} \}$}} are linearly independent,}}
then $\boldsymbol{v}'$ could be
rewritten as
{{$
\boldsymbol{v}' =  \mathsf{a}'_{j,s}\mathsf{\boldsymbol{z}}'_{1,j} +
\mathsf{b}'_{k,t}
\mathsf{\boldsymbol{z}}'_{2,k} + \bar{\mathsf{\boldsymbol{z}}}' $}},
where for some $s\in{1, \ldots, \abs{\mathcal{A}_{j}}}$ and
$t\in{1, \ldots, \abs{\mathcal{B}_{k}}}$ satisfying
\begin{equation}
\mathsf{b}_{k,\hat{t}}-\mathsf{b}_{k,t}=\frac{c_{2}}{c_{1}}\left(\mathsf{a}_{j,\hat{s}}-\mathsf{a}_{j,s}\right).
 \label{connect_condition}
\end{equation}
We will show that \eqref{connect_condition} is always not true. Since $\frac{c_{2}}{c_{1}}\mathsf{a}_{j,\hat{s}}$
has the smallest real part among $\frac{c_{2}}{c_{1}}\mathsf{a}_{j,1},\frac{c_{2}}{c_{1}}\mathsf{a}_{j,2},\ldots,\frac{c_{2}}{c_{1}}\mathsf{a}_{j,\left|\mathcal{A}_{j}\right|}$,
for $s\notin\mathcal{S}
 $, i.e., $\Re\left\{ \frac{c_{2}}{c_{1}}\mathsf{a}_{j,\hat{s}}\right\} \neq\Re\left\{ \frac{c_{2}}{c_{1}}\mathsf{a}_{j,s}\right\}$, we have
{{$
\Re\left\{ \frac{c_{2}}{c_{1}}\left(\mathsf{a}_{j,\hat{s}}-\mathsf{a}_{j,s}\right)\right\} < 0$}}. On the other hand, $\mathsf{b}_{k,\hat{t}}$ has the largest real part among $\mathcal{B}_{k}$, which indicates $\Re\left\{ \mathsf{b}_{k,\hat{t}}-\mathsf{b}_{k,t}\right\} \geq0$.
Hence, \eqref{connect_condition} is not true for $\Re\left\{ \frac{c_{2}}{c_{1}}\mathsf{a}_{j,\hat{s}}\right\} \neq\Re\left\{ \frac{c_{2}}{c_{1}}\mathsf{a}_{j,s}\right\}$.
For some $s\in\mathcal{S}$ that $\Re\left\{ \frac{c_{2}}{c_{1}}\mathsf{a}_{j,\hat{s}}\right\} =\Re\left\{ \frac{c_{2}}{c_{1}}\mathsf{a}_{j,s}\right\}$,
notice that $\frac{c_{2}}{c_{1}}\mathsf{a}_{j,\hat{s}}$ also has the smallest imaginary part among those whose real part equal to $\Re\left\{ \frac{c_{2}}{c_{1}}\mathsf{a}_{j,\hat{s}}\right\}$, we thus have $\frac{c_{2}}{c_{1}}\left(\mathsf{a}_{j,\hat{s}}-\mathsf{a}_{j,s}\right)=\mathsf{i}\Im\left\{ \frac{c_{2}}{c_{1}}\mathsf{a}_{j,\hat{s}}-\frac{c_{2}}{c_{1}}\mathsf{a}_{j,s}\right\} $, where
{{\[
\Im\left\{ \frac{c_{2}}{c_{1}}\mathsf{a}_{j,\hat{s}}-\frac{c_{2}}{c_{1}}\mathsf{a}_{j,s}\right\} <0
\]}}for $\Re\left\{ \frac{c_{2}}{c_{1}}\mathsf{a}_{j,\hat{s}}\right\} =\Re\left\{ \frac{c_{2}}{c_{1}}\mathsf{a}_{j,s}\right\}$.
In such case, $\mathsf{b}_{k,t}$ satisfying (\ref{connect_condition}) have the same real part as $\mathsf{b}_{k,\hat{t}}$.
$\mathsf{b}_{k,\hat{t}}$ also has the largest imaginary part among those whose real part equal to $\mathsf{b}_{k,\hat{t}}$,
which indicates $\mathsf{b}_{k,\hat{t}}-\mathsf{b}_{k,t}=\mathsf{i}\Im\left\{ \mathsf{b}_{k,\hat{t}}-\mathsf{b}_{k,t}\right\} $, and
{{\[
\Im\left\{ \mathsf{b}_{k,\hat{t}}-\mathsf{b}_{k,t}\right\} >0.
\]}} As a result, \eqref{connect_condition} neither can be established for $\Re\left\{ \frac{c_{2}}{c_{1}}\mathsf{a}_{j,\hat{s}}\right\} =\Re\left\{ \frac{c_{2}}{c_{1}}\mathsf{a}_{j,s}\right\}$. We obtain that at least $\boldsymbol{v} = \mathsf{a}_{j,\hat{s}}\mathsf{\boldsymbol{z}}'_{1,j} +
\mathsf{b}_{k,\hat{t}}
\mathsf{\boldsymbol{z}}'_{2,k} + \bar{\mathsf{\boldsymbol{z}}}' $ cannot be covered by $c_{1,j}\mathsf{\boldsymbol{z}}'_{1,j} +
c_{2,k}\mathsf{\boldsymbol{z}}'_{2,k}$.
The same argument also
generalizes for each vector of the form
$\sum_{j=1}^{N_{L}}c_{1,j}\mathsf{\boldsymbol{z}}'_{1,j} +
\sum_{k=1}^{N_{M}}c_{2,k}\mathsf{\boldsymbol{z}}'_{2,k}$.  The
proposition is thus proved.

\end{IEEEproof}

\section{Proof of Proposition 2}
We first illustrate in Appendix B.A  {{how}} 
the empirical distribution of the BS's observations
is determined by attack actions, channels and added noise.
Then, in Appendix B.B, we show that $\Phi_{\triangle {F}_{\widetilde{\ensuremath{\mathbf{z}}}}}\left( \boldsymbol{\omega}\right)$
asymptotically approaches to the product of $\Phi_{\triangle {F}_{\widetilde{\ensuremath{\mathbf{z}}}'}}\left( \boldsymbol{\omega}\right)$
and $\Phi_{F_{\mathbf{n}}} \left(\boldsymbol{\omega}\right)$. Finally, in  in Appendix B.C, we complete the proof by
using FFT and IFFT to get $\Phi_{\triangle {F}_{\widetilde{\ensuremath{\mathbf{z}}}}}\left( \boldsymbol{\omega}\right)$
and $\triangle \widehat{F}_{\widetilde{\ensuremath{\mathbf{z}}}'}$, respectively.

\subsection{Preliminary}
For $k=1,\ldots, \abs{\widetilde{\mathcal{Z}}}$ and $j=1,\ldots, \abs{\widetilde{\mathcal{Z'}}}$,
we define
\begin{equation}
f_{\mathbf{z}\left|\mathbf{z}'\right.}\left(\boldsymbol{u}\left|\widetilde{\boldsymbol{v}}_{j}\right.\right)=\frac{\sqrt{M}}{\sigma\sqrt{\pi}}\exp\left\{ -\frac{\abs{\boldsymbol{u}-\widetilde{\boldsymbol{v}}_{j}}^{2}}{\frac{\sigma^{2}}{2M}}\right\}
\end{equation}and
\begin{equation}\label{con_z'}
P_{\widetilde{\mathbf{z}}\left|\mathbf{z}'\right.}\left(\widetilde{\boldsymbol{u}}_{k}\left|\widetilde{\boldsymbol{v}}_{j}\right.\right)=\int_{\boldsymbol{u}\in\mathcal{B}\left(\widetilde{\boldsymbol{u}}_{k}\right)}f_{\mathbf{z}\left|\mathbf{z}'\right.}\left(\boldsymbol{u}\left|\widetilde{\boldsymbol{v}}_{j}\right.\right)d\boldsymbol{u},
\end{equation}where $\boldsymbol{u}$ and $\widetilde{\boldsymbol{u}}_{k}$
are $m_1\times 1$ vectors. We rewrite
$\boldsymbol{u}$ and $\widetilde{\boldsymbol{u}}_{k}$
as $\boldsymbol{u}=\left[u_{1},\ldots,u_{N_{L}+N_{M}}\right]^T$ and
$\widetilde{\boldsymbol{u}}_{k}=\left[\mathsf{u}_{k_1},\ldots,\mathsf{u}_{k_{N_{L}+N_{M}}}\right]^T$.
Notice $\widetilde{\boldsymbol{u}}_{k}\in \mathcal{Z}$, hence the
$i$th
element of $\widetilde{\boldsymbol{u}}_{k}$ is $\mathsf{u}_{k_i}\in\left\{ \mathsf{\widetilde{u}}_{1},\ldots,\mathsf{\widetilde{u}}_{m_{1}}\right\} $,
$i=1,\ldots,N_{L}+N_{M}$. Then $\mathcal{B}\left(\widetilde{\boldsymbol{u}}_{k}\right)$ in (\ref{con_z'})
is given by  $\mathcal{B}\left(\widetilde{\boldsymbol{u}}_{k}\right) =
\left\{ \boldsymbol{u}:u_{i}\in\mathcal{B}\left(\mathsf{u}_{k_i}\right),i=1,\ldots,N_{T}+N_{M}\right\} $.

\begin{lemma}\label{lem2}
For $k=1,\ldots, \abs{\widetilde{\mathcal{Z}}}$,
$\triangle F_{\widetilde{\ensuremath{\mathbf{z}}}^{n}}\left(\widetilde{\boldsymbol{u}}_{k}\right)\rightarrow\sum_{j=1}^{\left|\widetilde{\mathcal{Z}}'\right|}\triangle F_{\widetilde{\ensuremath{\mathbf{z}'}}^{n}}\left(\widetilde{\boldsymbol{v}}_{j}\right)P_{\widetilde{\mathbf{z}}\left|\mathbf{z}'\right.}\left(\widetilde{\boldsymbol{u}}_{k}\left|\widetilde{\boldsymbol{v}}_{j}\right.\right)$
in probability as $n$ and $m_2$ approach to infinity with $n = \Omega
\left( m_2^{2(N_L+N_M)} \right)$. To be more
precise, for sufficient large $m_2$ and $n$ with $n = \Omega
\left( m_2^{2(N_L+N_M)} \right)$, we  have {{}{
\begin{equation}
\Pr\left\{ \left|\triangle
   F_{\widetilde{\ensuremath{\mathbf{z}}}}\left(\widetilde{\boldsymbol{u}}_{k}\right)-\sum_{j=1}^{\left|\widetilde{\mathcal{Z}}'\right|}\triangle
   F_{\widetilde{\ensuremath{\mathbf{z}}}'}\left(\widetilde{\boldsymbol{v}}_{j}\right)
   P_{\widetilde{\mathbf{z}}\left|\mathbf{z}'\right.}\left(\widetilde{\boldsymbol{u}}_{k}\left|\widetilde{\boldsymbol{v}}_{j}\right.\right)\right|
   \geq \mu\right\} \nonumber \\
 \leq\frac{1}{\mu^{2}}\mathcal{O}\left( \delta_2^{N_L+N_M}\right).
\end{equation}
}} \end{lemma}
Please see Appendix C for the proof of Lemma \ref{lem2}.
From Lemma \ref{lem2}, it is easy to get the following
corollary:
\begin{cor}
\label{cor1}
By choosing $m_{2}=\Omega \left(m_{1}^{3}\right)$, we have
$\sum_{k=1}^{\left|\widetilde{\mathcal{Z}}\right|}\left|\triangle
  F_{\widetilde{\ensuremath{\mathbf{z}}}}\left(\widetilde{\boldsymbol{u}}_{k}\right)-\sum_{j=1}^{\left|\widetilde{\mathcal{Z}}'\right|}\triangle
  F_{\widetilde{\ensuremath{\mathbf{z}}}'}\left(\widetilde{\boldsymbol{v}}_{j}\right)P_{\widetilde{\mathbf{z}}\left|\mathbf{z}'\right.}\left(\widetilde{\boldsymbol{u}}_{k}\left|\widetilde{\boldsymbol{v}}_{j}\right.\right)\right|\rightarrow0$
in probability.
\end{cor}
\begin{IEEEproof}
{{\begin{align*}
&\Pr\left\{
  \sum_{k=1}^{\left|\widetilde{\mathcal{Z}}\right|}\left|\triangle
  F_{\widetilde{\ensuremath{\mathbf{z}}}}\left(\widetilde{\boldsymbol{u}}_{k}\right)-\sum_{j=1}^{\left|\widetilde{\mathcal{Z}}'\right|}\triangle
  F_{\widetilde{\ensuremath{\mathbf{z}}}'}\left(\widetilde{\boldsymbol{v}}_{j}\right)P_{\widetilde{\mathbf{z}}\left|\mathbf{z}'\right.}\left(\widetilde{\boldsymbol{u}}_{k}\left|\widetilde{\boldsymbol{v}}_{j}\right.\right)\right|\geq\mu_{1}\right\}\\& <\sum_{k=1}^{\left|\widetilde{\mathcal{Z}}\right|}\Pr\left\{
  \left|\triangle
  F_{\widetilde{\ensuremath{\mathbf{z}}}}\left(\widetilde{\boldsymbol{u}}_{k}\right)-\sum_{j=1}^{\left|\widetilde{\mathcal{Z}}'\right|}\triangle
  F_{\widetilde{\ensuremath{\mathbf{z}}}'}\left(\widetilde{\boldsymbol{v}}_{j}\right)P_{\widetilde{\mathbf{z}}\left|\mathbf{z}'\right.}\left(\widetilde{\boldsymbol{u}}_{k}\left|\widetilde{\boldsymbol{v}}_{j}\right.\right)\right|\geq\frac{\mu_{1}}{\left|\widetilde{\mathcal{Z}}\right|}\right\} \\
&<\frac{\left|\widetilde{\mathcal{Z}}\right|^{3}}{\mu^{2}}\mathcal{O}\left(
  \delta_2^{N_L+N_M}\right),
\end{align*}}}where the last inequality follows Lemma \ref{lem2}.
Recall that $\abs{\widetilde{\mathcal{Z}}} = m_{1}^{N_{L}+N_{M}}$
 and
$m_2 =\frac{2\alpha_2}{\delta_2}+1 $.
 Therefore,
choosing $m_{2}=\Omega \left(m_{1}^{3}\right)$ will prove the corollary.
\end{IEEEproof}
\subsection{Convergences of characteristic functions}
To obtain the proof of Proposition 2, we employ an auxiliary random variable $\mathbf{u}$, which
is specified by $\mathbf{u}=\mathbf{u}'+\mathbf{n}$, where the pmf of $\mathbf{u}'$
is given by
$P_{\mathbf{u}'}\left(\widetilde{\boldsymbol{v}}_{j}\right)=\triangle
F_{\widetilde{\ensuremath{\mathbf{z}'}}}\left(\widetilde{\boldsymbol{v}}_{j}\right)$
for
$j=1,\ldots, \left|\widetilde{\mathcal{Z}}'\right|$, and
$\mathbf{u}'$ is independent of $\mathbf{n}$. Hence, the pdf of $\mathbf{u}$ is given by
{{$
f_{\mathbf{u}}\left(\boldsymbol
  u\right)=\sum_{j=1}^{\left|\widetilde{\mathcal{Z}}'\right|}\triangle
F_{\widetilde{\ensuremath{\mathbf{z}}}'}\left(\widetilde{\boldsymbol{v}}_{j}\right)f_{\mathbf{z}\left|\mathbf{z}'\right.}\left(\boldsymbol u\left|\widetilde{\boldsymbol{v}}_{j}\right.\right)$}}.
According to the fact that $\mathbf{u}'$ is independent of $\mathbf{n}$,
we have
\begin{equation}
\Phi\left(f_{\mathbf{u}}\right)=\Phi\left(\triangle
  F_{\widetilde{\ensuremath{\mathbf{z}}}'}\right)\Phi\left(f_{\mathbf{n}}\right),\label{p3}
\end{equation}
where
{{$
\Phi\left(\triangle
  F_{\widetilde{\ensuremath{\mathbf{z}}}'}\right)=\sum_{j=1}^{\left|\widetilde{\mathcal{Z}}'\right|}\triangle
F_{\widetilde{\ensuremath{\mathbf{z}}}'}\left(\widetilde{\boldsymbol{v}}_{j}\right)\exp\left\{ -\mathsf{i}\left[\Re\left\{ \widetilde{\boldsymbol{v}}_{j}\right\} ,\Im\left\{ \widetilde{\boldsymbol{v}}_{j}\right\} \right]^{T}\boldsymbol{\omega}\right\}
$}}and
{{$
\Phi\left(f_{\mathbf{n}}\right)=\exp\left\{ -{\abs{ \boldsymbol{\omega}}^{2}}{\frac{\sigma^{2}}{4M}}\right\}
$}}.
On the other hand, according to the expression of $f_{\mathbf{u}}\left(\boldsymbol u\right)$,
$\Phi\left(f_{\mathbf{u}}\right)$ is given by
{{\begin{align}
 \Phi\left(f_{\mathbf{u}}\left(\boldsymbol u\right)\right)=\hspace{10pt} \sum_{j=1}^{\left|\widetilde{\mathcal{Z}}'\right|}\triangle
   F_{\widetilde{\ensuremath{\mathbf{z}}}'}\left(\widetilde{\boldsymbol{v}}_{j}\right)\int_{-\infty}^{+\infty}f_{\mathbf{z}\left|\mathbf{z}'\right.}\left(\boldsymbol
   u\left|\widetilde{\boldsymbol{v}}_{j}\right.\right)\exp\left\{ -\mathsf{i}\left[\Re\left\{ \boldsymbol{u}\right\} ,\Im\left\{ \boldsymbol{u}\right\} \right]^{T}\boldsymbol{\omega}\right\}
   d\boldsymbol u.
\end{align}}}
Hence, we have {{}{
\begin{align*}
&\bigg| \int_{-\infty}^{+\infty}f_{\mathbf{z}\left|\mathbf{z}'\right.}\left(\boldsymbol u\left|\widetilde{\boldsymbol{v}}_{j}\right.\right)\exp\left\{ -\mathsf{i}\left[\Re\left\{ \boldsymbol{u}\right\} ,\Im\left\{ \boldsymbol{u}\right\} \right]^{T}\boldsymbol{\omega}\right\}  d\boldsymbol u -\\&\hspace{105pt}\sum_{k=1}^{\left|\widetilde{\mathcal{Z}}\right|}P_{\widetilde{\mathbf{z}}\left|\mathbf{z}'\right.}\left(\widetilde{\boldsymbol{u}}_{k}\left|\widetilde{\boldsymbol{v}}_{j}\right.\right)\exp\left\{ -\mathsf{i}\left[\Re\left\{ \widetilde{\boldsymbol{u}}_{k}\right\} ,\Im\left\{ \widetilde{\boldsymbol{u}}_{k}\right\} \right]^{T}\boldsymbol{\omega}\right\}  \bigg|\\&<\gamma_{m_{1}},
\end{align*}
}}where $\lim_{m_{1}\rightarrow\infty}\gamma_{m_{1}}=0$.
Then, we have
{{\begin{equation}
\big|\sum_{k=1}^{\left|\widetilde{\mathcal{Z}}\right|}\sum_{j=1}^{\left|\widetilde{\mathcal{Z}}'\right|}\triangle
           F_{\widetilde{\ensuremath{\mathbf{z}}}'}\left(\widetilde{\boldsymbol{v}}_{j}\right)P_{\widetilde{\mathbf{z}}\left|\mathbf{z}'\right.}\left(\widetilde{\boldsymbol{u}}_{k}\left|\widetilde{\boldsymbol{v}}_{j}\right.\right)\exp\left\{ -\mathsf{i}\left[\Re\left\{ \widetilde{\boldsymbol{u}}_{k}\right\} ,\Im\left\{ \widetilde{\boldsymbol{u}}_{k}\right\} \right]^{T}\boldsymbol{\omega}\right\}-\Phi\left(f_{\mathbf{u}}\left(\boldsymbol{u}\right)\right) \big|
\leq\gamma_{m_{1}} \label{p2}
        \end{equation}}}
    Based on Corollary 1, and the fact that $\Phi\left(\cdot\right)$ is an
    orthogonal transformation,
    we get for arbitrary frequency $\boldsymbol{\omega}$,
    there has
    {{\begin{align}\label{p1}
  \nonumber         &   \sum_{k=1}^{\left|\widetilde{\mathcal{Z}}\right|}\triangle F_{\widetilde{\ensuremath{\mathbf{z}}}}\left(\widetilde{\boldsymbol{u}}_{k}\right)\exp\left\{ -\mathsf{i}\left[\Re\left\{ \widetilde{\boldsymbol{u}}_{k}\right\} ,\Im\left\{ \widetilde{\boldsymbol{u}}_{k}\right\} \right]^{T}\boldsymbol{\omega}\right\}  \rightarrow    \\&\hspace{75pt}            \sum_{k=1}^{\left|\widetilde{\mathcal{Z}}\right|}\sum_{j=1}^{\left|\widetilde{\mathcal{Z}}'\right|}\triangle F_{\widetilde{\ensuremath{\mathbf{z}}}'}\left(\widetilde{\boldsymbol{v}}_{j}\right)P_{\widetilde{\mathbf{z}}\left|\mathbf{z}'\right.}\left(\widetilde{\boldsymbol{u}}_{k}\left|\widetilde{\boldsymbol{v}}_{j}\right.\right)\exp\left\{ -\mathsf{i}\left[\Re\left\{ \widetilde{\boldsymbol{u}}_{k}\right\} ,\Im\left\{ \widetilde{\boldsymbol{u}}_{k}\right\} \right]^{T}\boldsymbol{\omega}\right\}
\end{align}}}in probability as ${m_{1}\rightarrow\infty}$, ${m_{2}\rightarrow\infty}$, and ${n\rightarrow\infty}$.
This convergence follows the fact that
{{\begin{align}\label{ineq1}
&\nonumber\big|\sum_{k=1}^{\left|\widetilde{\mathcal{Z}}\right|}\big\{
  \sum_{j=1}^{\left|\widetilde{\mathcal{Z}}'\right|}\triangle
  F_{\widetilde{\ensuremath{\mathbf{z}}}'}\left(\widetilde{\boldsymbol{v}}_{j}\right)P_{\widetilde{\mathbf{z}}\left|\mathbf{z}'\right.}\left(\widetilde{\boldsymbol{u}}_{k}\left|\widetilde{\boldsymbol{v}}_{j}\right.\right)-\sum_{k=1}^{\left|\widetilde{\mathcal{Z}}\right|}\triangle F_{\widetilde{\ensuremath{\mathbf{z}}}}\left(\widetilde{\boldsymbol{u}}_{k}\right)\big\}\exp\left\{ -\mathsf{i}\left[\Re\left\{ \widetilde{\boldsymbol{u}}_{k}\right\} ,\Im\left\{ \widetilde{\boldsymbol{u}}_{k}\right\} \right]^{T}\boldsymbol{\omega}\right\}  \big|
\\&<\sum_{k=1}^{\left|\widetilde{\mathcal{Z}}\right|}\left|\sum_{j=1}^{\left|\widetilde{\mathcal{Z}}'\right|}\triangle
                                                                                                                                                                                                                                                                                                                                                                                                                                                                                                         F_{\widetilde{\ensuremath{\mathbf{z}}}'}\left(\widetilde{\boldsymbol{v}}_{j}\right)P_{\widetilde{\mathbf{z}}\left|\mathbf{z}'\right.}\left(\widetilde{\boldsymbol{u}}_{k}\left|\widetilde{\boldsymbol{v}}_{j}\right.\right)-\sum_{k=1}^{\left|\widetilde{\mathcal{Z}}\right|}\triangle F_{\widetilde{\ensuremath{\mathbf{z}}}}\left(\widetilde{\boldsymbol{u}}_{k}\right)\right|.
\end{align}}}
Apply Corollary 1 to (\ref{ineq1}), we can obtain the convergence given by (\ref{p1}).

Further, we have
{{\begin{align}\label{estCon}
\nonumber&\Pr\big\{
  \big|\sum_{k=1}^{\left|\widetilde{\mathcal{Z}}\right|}\triangle
  F_{\widetilde{\ensuremath{\mathbf{z}}}}\left(\widetilde{\boldsymbol{u}}_{k}\right)\exp\left\{ -\mathsf{i}\left[\Re\left\{ \widetilde{\boldsymbol{u}}_{k}\right\} ,\Im\left\{ \widetilde{\boldsymbol{u}}_{k}\right\} \right]^{T}\boldsymbol{\omega}\right\}
  -\\&\hspace{65pt}\exp\left\{ -\frac{\sigma^{2}}{4M}\abs{
                \boldsymbol{\omega}} ^{2}\right\}
                \sum_{j=1}^{\left|\widetilde{\mathcal{Z}}'\right|}\triangle
                F_{\widetilde{\ensuremath{\mathbf{z}}}'}\left(\widetilde{\boldsymbol{v}}_{j}\right)\exp\left\{ -\mathsf{i}\left[\Re\left\{ \widetilde{\boldsymbol{v}}_{j}\right\} ,\Im\left\{ \widetilde{\boldsymbol{v}}_{j}\right\} \right]^{T}\boldsymbol{\omega}\right\} \big|\geq\mu_{1}\big\}\\
\nonumber&\leq\Pr\big\{
           \big|\sum_{k=1}^{\left|\widetilde{\mathcal{Z}}\right|}\sum_{j=1}^{\left|\widetilde{\mathcal{Z}}'\right|}\triangle
           F_{\widetilde{\ensuremath{\mathbf{z}}}'}\left(\widetilde{\boldsymbol{v}}_{j}\right)P_{\widetilde{\mathbf{z}}\left|\mathbf{z}'\right.}\left(\widetilde{\boldsymbol{u}}_{k}\left|\widetilde{\boldsymbol{v}}_{j}\right.\right)\exp\left\{ -\mathsf{i}\left[\Re\left\{ \widetilde{\boldsymbol{u}}_{k}\right\} ,\Im\left\{ \widetilde{\boldsymbol{u}}_{k}\right\} \right]^{T}\boldsymbol{\omega}\right\}
           -\\\nonumber&\hspace{100pt}\exp\left\{ -\frac{\sigma^{2}}{4M}\abs{
                         \boldsymbol{\omega}} ^{2}\right\}
                         \sum_{j=1}^{\left|\widetilde{\mathcal{Z}}'\right|}\triangle
                         F_{\widetilde{\ensuremath{\mathbf{z}}}'}\left(\widetilde{\boldsymbol{v}}_{j}\right)\exp\left\{ -\mathsf{i}\left[\Re\left\{ \widetilde{\boldsymbol{v}}_{j}\right\} ,\Im\left\{ \widetilde{\boldsymbol{v}}_{j}\right\} \right]^{T}\boldsymbol{\omega}\right\}  \big|\geq\frac{\mu_{1}}{2}\big\} \\\nonumber&
+\Pr\big\{
                                                                                                                                                                                                                                                                                                                                                                                                                                                                                                                                                                                                                                                                                                                       \big|\sum_{k=1}^{\left|\widetilde{\mathcal{Z}}\right|}\triangle F_{\widetilde{\ensuremath{\mathbf{z}}}}\left(\widetilde{\boldsymbol{u}}_{k}\right)\exp\left\{ -\mathsf{i}\left[\Re\left\{ \widetilde{\boldsymbol{u}}_{k}\right\} ,\Im\left\{ \widetilde{\boldsymbol{u}}_{k}\right\} \right]^{T}\boldsymbol{\omega}\right\}  -\\\nonumber&\hspace{55pt}\sum_{k=1}^{\left|\widetilde{\mathcal{Z}}\right|}\sum_{j=1}^{\left|\widetilde{\mathcal{Z}}'\right|}\triangle F_{\widetilde{\ensuremath{\mathbf{z}}}'}\left(\widetilde{\boldsymbol{v}}_{j}\right)P_{\widetilde{\mathbf{z}}\left|\mathbf{z}'\right.}\left(\widetilde{\boldsymbol{u}}_{k}\left|\widetilde{\boldsymbol{v}}_{j}\right.\right)\exp\left\{ -\mathsf{i}\left[\Re\left\{ \widetilde{\boldsymbol{u}}_{k}\right\} ,\Im\left\{ \widetilde{\boldsymbol{u}}_{k}\right\} \right]^{T}\boldsymbol{\omega}\right\}  \big|\geq\frac{\mu_{1}}{2}\big\}.
 \end{align}}}Based on (\ref{p2}) and (\ref{p1}), the two items on the right side of (\ref{estCon}) approach to zero.
We thus
finally get {{}{
\begin{equation}
{\frac{\sum_{k=1}^{\left|\widetilde{\mathcal{Z}}\right|}\triangle
    F_{\widetilde{\ensuremath{\mathbf{z}}}}\left(\widetilde{\boldsymbol{u}}_{k}\right)\exp\left\{ -\mathsf{i}\left[\Re\left\{ \widetilde{\boldsymbol{u}}_{k}\right\} ,\Im\left\{ \widetilde{\boldsymbol{u}}_{k}\right\} \right]^{T}\boldsymbol{\omega}\right\}
 }{\exp\left\{ -\frac{\sigma^{2}}{4M}\abs{
        \boldsymbol{\omega}} ^{2}\right\}
  }}\rightarrow\sum_{j=1}^{\left|\widetilde{\mathcal{Z}}'\right|}\triangle
F_{\widetilde{\ensuremath{\mathbf{z}}}'}\left(\widetilde{\boldsymbol{v}}_{j}\right)\exp\left\{ -\mathsf{i}\left[\Re\left\{ \widetilde{\boldsymbol{v}}_{j}\right\} ,\Im\left\{ \widetilde{\boldsymbol{v}}_{j}\right\} \right]^{T}\boldsymbol{\omega}\right\} \label{estimation_A}
\end{equation}
}}in probability as ${m_{1}\rightarrow\infty}$, ${m_{2}\rightarrow\infty}$, and ${n\rightarrow\infty}$.
The left side of (\ref{estimation_A}) is what the BS observes,
the right side of (\ref{estimation_A}) is what the BS intends to
estimate. Hence, we may estimate $\triangle
F_{\widetilde{\ensuremath{\mathbf{z}}}'}\left(\widetilde{\boldsymbol{v}}_{j}\right)$
according to (\ref{estimation_A}).

On the other hand, notice that
{{\begin{align}\label{y_c}
\nonumber&\sum_{j=1}^{\left|\mathcal{Z}\right|}\triangle F_{\widetilde{\ensuremath{\mathbf{z}}}}\left(\widetilde{\boldsymbol{u}}_{j}\right)\exp\left\{ -\mathsf{i}\left[\Re\left\{ \widetilde{\boldsymbol{u}}_{k}\right\} ,\Im\left\{ \widetilde{\boldsymbol{u}}_{k}\right\} \right]^{T}\boldsymbol{\omega}\right\}  =\exp\left\{ \mathsf{i}\alpha_{1}\sum_{t=1}^{2N_{L}+2N_{M}}\omega_{t}\right\} F_{1}\left(\mathbf{\boldsymbol{\omega}}\right),
\end{align}}}where 
\begin{align*}
&F_{1}\left(\mathbf{\boldsymbol{\omega}}\right)=\\&\sum_{j_{1}=1}^{m_{1}}\cdots\sum_{j_{2N_{L}+2N_{M}}=1}^{m_{1}}\triangle F_{\widetilde{\ensuremath{\mathbf{z}}}}\left(\left[\mathsf{u}_{j_{1}},\ldots,\mathsf{u}_{j_{N_{L}+N_{M}}}\right]\right)\exp\left\{ -\mathsf{i}\sum_{t=1}^{2N_{L}+2N_{M}}\delta_1\left(j_{t}-1\right)\omega_{t}\right\}. 
\end{align*}
And
{{\begin{equation}\label{z_c}
\sum_{j=1}^{\left|\mathcal{Z}'\right|}\triangle
F_{\widetilde{\ensuremath{\mathbf{z}}}'}\left(\widetilde{\boldsymbol{v}}_{j}\right)\exp\left\{ -\mathsf{i}\left[\Re\left\{ \widetilde{\boldsymbol{v}}_{j}\right\} ,\Im\left\{ \widetilde{\boldsymbol{v}}_{j}\right\} \right]^{T}\boldsymbol{\omega}\right\} =\exp\left\{ \mathsf{i}\alpha_{2}\sum_{t=1}^{2N_{L}+2N_{M}}\omega_{t}\right\}  F_{2}\left(\boldsymbol{\omega}\right),
\end{equation}}}where {{\begin{equation}
                                                      F_{2}\left(\boldsymbol{\omega}\right)=\sum_{j_{1}=1}^{m_{2}-1}\cdots\sum_{j_{2N_{L}+2N_{M}}=1}^{m_{2}-1}\triangle F_{\widetilde{\ensuremath{\mathbf{z}}}'}\left(\left[\mathsf{v}_{j_{1}},\ldots,\mathsf{v}_{j_{N_{L}+N_{M}}}\right]\right)\exp\left\{ -\mathsf{i}\sum_{t=1}^{2N_{L}+2N_{M}}\delta_2\left(j_{t}-1\right)\omega_{t}\right\}\end{equation}}}is obtained based on the assumption that $\Re\left\{ \mathcal{Z}\right\}\subseteq\left[-\alpha_{1},\alpha_{1}-\delta_2\right]^{N_{M}+N_{L}}$,
$\Im\left\{ \mathcal{Z}\right\} \subseteq\left[-\alpha_{1},\alpha_{1}-\delta_2\right]^{N_{M}+N_{L}}$.
To be more precise, for $t=1,\ldots,2N_{L}+2N_{M}$, if $j_t=m_2$,   $\triangle F_{\widetilde{\ensuremath{\mathbf{z}}}'}\left(\left[\mathsf{v}_{j_{1}},\ldots,\mathsf{v}_{j_{t}},\ldots\mathsf{v}_{j_{N_{L}+N_{M}}}\right]\right)=0$. These points in the boundary of quantization range are trivial to estimation.

Then, (\ref{estimation_A}) becomes {{}{
\begin{equation}
\underbrace{\frac{\exp\left\{ \mathsf{i}\alpha_{1}\sum_{t=1}^{2N_{L}+2N_{M}}\omega_{t}\right\} }{\exp\left\{ \mathsf{i}\alpha_{2}\sum_{t=1}^{2N_{L}+2N_{M}}\omega_{t}\right\} }\frac{F_{1}\left(\boldsymbol{\omega}\right)}{\exp\left\{ -\frac{\sigma^{2}}{4M}\abs{ \boldsymbol{\omega}}^{2}\right\} }}_{F'_{1}\left(\boldsymbol{\omega}\right)}\rightarrow F_{2}\left(\boldsymbol{\omega}\right)\label{reshape_estimation}
\end{equation}
}} 
\subsection{Complete the proof using FFT and IFFT}
Let us choose $\boldsymbol{\omega}$ from $\mathcal{W}=\left\{
  \boldsymbol{\omega}\left|\omega_{1},\ldots,\omega_{2N_L+2N_M}=0,\,\frac{2\pi}{m_2\delta_2}\,,\ldots,\frac{2\pi\left(m_2-2\right)}{(m_2-1)\delta_2}\right.\right\}
$.
Recall that, we set $\alpha_{1}=\alpha_{2}$,
then $F'_{1}\left(\boldsymbol{\omega}\right)$ in (\ref{reshape_estimation}) becomes
\begin{equation}
F'_{1}\left(\boldsymbol{\omega}\right)=\frac{F_{1}\left(\boldsymbol{\omega}\right)}{\exp\left\{ -\frac{\sigma^{2}}{4M}\abs{\boldsymbol{\omega}}^{2}\right\} }.
\end{equation}
By setting $\alpha_{1}=\alpha_{2}$, $2\alpha_{1}=(m_1-1)\delta_1$, $2\alpha_{2}=(m_2-1)\delta_2$, and $\frac{m_{2}-1}{m_{1}-1}
 $ is  integer,  $F_{1}\left(\boldsymbol{\omega}\right)$ over $\boldsymbol{\omega}\in\mathcal{W}$
 could be reshaped as
 {{\begin{align}\label{fftform}
& F_{1}\left({\mathbf{\boldsymbol{\omega}}}\right)=\sum_{j_{1}=1}^{m_{1}}\cdots\sum_{j_{N_{L}+N_{M}}=1}^{m_{1}}\triangle F_{\widetilde{\ensuremath{\mathbf{z}}}}\left(\left[\mathsf{u}_{j_{1}},\ldots,\mathsf{u}_{j_{N_{L}+N_{M}}}\right]\right)\exp\left\{ -\mathsf{i}\sum_{t=1}^{2N_{L}+2N_{M}}\delta_{1}\left(j_{t}-1\right)\frac{2\pi k_{t}}{(m_{2}-1)\delta_{2}}\right\} \\\nonumber&=\sum_{j_{1}=1}^{m_{1}}\cdots\sum_{j_{2N_{L}+2N_{M}}=1}^{m_{1}}\triangle F_{\widetilde{\ensuremath{\mathbf{z}}}}\left(\left[\mathsf{u}_{j_{1}},\ldots,\mathsf{u}_{j_{N_{L}+N_{M}}}\right]\right)\exp\left\{ -\mathsf{i}\sum_{t=1}^{2N_{L}+2N_{M}}\left(j_{t}-1\right)\frac{2\pi k_{t}}{(m_{1}-1)}\right\},
 \end{align}}}where {{$\mathbf{\boldsymbol{\omega}}=\left[\omega_{1},\ldots,\omega_{2N_{L}+2N_{M}}\right]$}},
 the first equation is based on the fact that $\boldsymbol{\omega}\in\mathcal{W}$ indicates for $t=1,\ldots,N_{L}+N_{M}$,
$\omega_{t}=\frac{2\pi k_{t}}{(m_{2}-1)\delta_{2}}$, $ k_{t}=0,\ldots m_{2}-2$. The second equation is based on setup
that $\alpha_{1}=\alpha_{2}$, $2\alpha_{1}=(m_1-1)\delta_1$, $2\alpha_{2}=(m_2-1)\delta_2$.
The second equation of (\ref{fftform}) is just $(2N_L+2N_M)$ dimension FFT expression of $\triangle F_{\widetilde{\ensuremath{\mathbf{z}}}}\left(\left[\mathsf{u}_{j_{1}},\ldots,\mathsf{u}_{j_{N_{L}+N_{M}}}\right]\right)$ over $m_1-1$ points. Hence, $F'_{1}\left(\boldsymbol{\omega}\right)$
could be obtained by applying FFT to $\triangle F_{\widetilde{\ensuremath{\mathbf{z}}}}\left(\left[\mathsf{u}_{j_{1}},\ldots,\mathsf{u}_{j_{N_{L}+N_{M}}}\right]\right)$.
Similarly, all values of ${F_{2}\left(\boldsymbol{\omega}\right)}$ over $\boldsymbol{\omega}\in\mathcal{W}$,
is the $(2N_L+2N_M)$-dimension FFT of $\triangle F_{\widetilde{\ensuremath{\mathbf{z}'}}}$ over $m_2-1$ points, i.e.,
{{\begin{equation}
\triangle{F}_{\widetilde{\ensuremath{\mathbf{z}}}'}=\Phi^{-1}\left(F_{2}\left(\boldsymbol{\omega}\right)_{\left|{\boldsymbol{\omega}\in\mathcal{W}}\right.}\right).\label{IFFT_e2}
\end{equation}}}Recall that $\Phi^{-1}\left(\cdot\right)$ denotes inverse FFT operation. $F_{2}\left(\boldsymbol{\omega}\right)_{\left|{\boldsymbol{\omega}\in\mathcal{W}}\right.}$
denotes the sequence that all values of $F_{2}\left(\boldsymbol{\omega}\right)$
over ${\boldsymbol{\omega}\in\mathcal{W}}$.



By applying inverse $(2N_L+2N_M)$-FFT transform to $F'_{1}\left(\boldsymbol{\omega}\right)$,
${\boldsymbol{\omega}\in\mathcal{W}}$, we complete the calculation of the estimator
 $\triangle\widehat{F}_{\widetilde{\ensuremath{\mathbf{z}}}'}$~(\ref{eq:hat}).
{{\begin{equation}
\triangle\widehat{F}_{\widetilde{\ensuremath{\mathbf{z}}}'}=\Phi^{-1}\left(F'_{1}\left(\boldsymbol{\omega}\right)_{\left|{\boldsymbol{\omega}\in\mathcal{W}}\right.}\right)\label{IFFT_e}
\end{equation}}}
where $F'_{1}\left(\boldsymbol{\omega}\right)_{\left|{\boldsymbol{\omega}\in\mathcal{W}}\right.}$
denotes the sequence that $F'_{1}\left(\boldsymbol{\omega}\right)$
over ${\boldsymbol{\omega}\in\mathcal{W}}$.

On the other hand, the convergence of (\ref{reshape_estimation}) indicates
$\left|F'_{1}\left(\boldsymbol{\omega}\right)-F_{2}\left(\boldsymbol{\omega}\right)\right|\leq\epsilon_{m_2}$ in probability,
$\epsilon_{m_2}\rightarrow0$, as $m_2$, $m_1$, and $n$ approach
to infinity. IFFT operation is orthogonal transformation, hence, we
have
\begin{equation}
\left|\Phi^{-1}\left(F'_{1}\left(\boldsymbol{\omega}\right)\right)-
  \Phi^{-1}
  \left(F_{2}\left(\boldsymbol{\omega}\right)\right)\right|\leq \epsilon_{m_2}
\end{equation}
According to (\ref{IFFT_e}), it is equivalent to
$\left|\triangle F_{\widetilde{\ensuremath{\mathbf{z}'}}}\left(\widetilde{\boldsymbol{v}}_{j}\right)-\triangle\widehat{F}_{\widetilde{\ensuremath{\mathbf{z}'}}}\left(\widetilde{\boldsymbol{v}}_{j}\right)\right|\leq\epsilon_{m_2}$
for $j=1,2,\ldots,\left|\widetilde{\mathcal{Z}}'\right|$, $\epsilon_{m_2}\rightarrow0$
as $m_1$ and $m_2$ approach to infinity.

\section{Proof of Lemma \ref{lem2}}

\begin{IEEEproof}
The goal of Lemma 1 is to prove
{{\small}{
\begin{align}
\nonumber&\Pr\left\{ \left|\triangle F_{\widetilde{\ensuremath{\mathbf{z}}}^{n}}\left(\widetilde{\boldsymbol{u}}_{k}\right)-\sum_{j=1}^{\left|\widetilde{\mathcal{Z}}'\right|}\triangle F_{\widetilde{\ensuremath{\mathbf{z}'}}^{n}}\left(\widetilde{\boldsymbol{v}}_{j}\right)P_{\widetilde{\mathbf{z}}\left|\mathbf{z}'\right.}\left(\widetilde{\boldsymbol{u}}_{k}\left|\widetilde{\boldsymbol{v}}_{j}\right.\right)\right|>\mu\right\} =\\&\Pr\big\{\left|\sum_{j=1}^{\left|\widetilde{\mathcal{Z}}'\right|}{\left(\frac{\sum_{i=1}^{n}\mathfrak{1}_{i}\left(\widetilde{\ensuremath{\mathbf{z}'}}_{i}=\widetilde{\boldsymbol{v}}_{j}\right)\mathfrak{1}_{i}\left(\mathbf{z}_{i}\in\mathcal{B}\left(\widetilde{\boldsymbol{u}}_{k}\right)\right)}{n}-\frac{\sum_{i=1}^{n}\mathfrak{1}_{i}\left(\widetilde{\ensuremath{\mathbf{z}'}}_{i}=\widetilde{\boldsymbol{v}}_{j}\right)}{n}P_{\widetilde{\mathbf{z}}\left|\mathbf{z}'\right.}\left(\widetilde{\boldsymbol{u}}_{k}\left|\widetilde{\boldsymbol{v}}_{j}\right.\right)\right)}\right|>\mu\big\}.\label{fupper3}
\end{align}
}}
Using the Chebyshev
inequality, we obtain
{{\small
\begin{align}
\nonumber&\Pr\left\{ \left|\sum_{j=1}^{\left|\widetilde{\mathcal{Z}}'\right|}{\underbrace{\left(\frac{\sum_{i=1}^{n}\mathfrak{1}_{i}\left(\widetilde{\ensuremath{\mathbf{z}'}}_{i}=\widetilde{\boldsymbol{v}}_{j}\right)\mathfrak{1}_{i}\left(\mathbf{z}_{i}\in\mathcal{B}\left(\widetilde{\boldsymbol{u}}_{k}\right)\right)}{n}-\frac{\sum_{i=1}^{n}\mathfrak{1}_{i}\left(\widetilde{\ensuremath{\mathbf{z}'}}_{i}=\widetilde{\boldsymbol{v}}_{j}\right)}{n}P_{\widetilde{\mathbf{z}}\left|\mathbf{z}'\right.}\left(\widetilde{\boldsymbol{u}}_{k}\left|\widetilde{\boldsymbol{v}}_{j}\right.\right)\right)}_{H_{j}}}\right|>\mu\right\} \\&\leq\frac{1}{\mu^{2}}\sum_{j=1}^{{\left|\widetilde{\mathcal{Z}}'\right|}}\sum_{j'=1}^{{\left|\widetilde{\mathcal{Z}}'\right|}}E\left(H_{j}H_{j'}\right),\label{fupper4}
\end{align}
}}
where $E\left(H_{j}H_{j'}\right)$
can be further extended in 
{{\tiny
\begin{align}\label{fupper5}
\nonumber & E\left(H_{j}H_{j'}\right)=\\&\frac{E\left(\sum_{i=1}^{n}\mathfrak{1}_{i}\left(\widetilde{\ensuremath{\mathbf{z}'}}_{i}=\widetilde{\boldsymbol{v}}_{j}\right)\left(\mathfrak{1}_{i}\left(\mathbf{z}_{i}\in\mathcal{B}\left(\widetilde{\boldsymbol{u}}_{k}\right)\right)-P_{\widetilde{\mathbf{z}}\left|\mathbf{z}'\right.}\left(\widetilde{\boldsymbol{u}}_{k}\left|\widetilde{\boldsymbol{v}}_{j}\right.\right)\right)\right)\left(\sum_{i=1}^{n}\mathfrak{1}_{i}\left(\widetilde{\ensuremath{\mathbf{z}'}}_{i}=\widetilde{\boldsymbol{v}}_{j'}\right)\left(\mathfrak{1}_{i}\left(\mathbf{z}_{i}\in\mathcal{B}\left(\widetilde{\boldsymbol{u}}_{k}\right)\right)-P_{\widetilde{\mathbf{z}}\left|\mathbf{z}'\right.}\left(\widetilde{\boldsymbol{u}}_{k}\left|\widetilde{\boldsymbol{v}}_{j'}\right.\right)\right)\right)}{n^{2}}\nonumber \\
 & <\frac{1}{n}+\nonumber \\&\frac{E\left(\sum_{i=1}^{n}\sum_{i\text{'}=1,i\neq i}^{n}\mathfrak{1}_{i}\left(\widetilde{\ensuremath{\mathbf{z}'}}_{i}=\widetilde{\boldsymbol{v}}_{j}\right)\left(\mathfrak{1}_{i}\left(\mathbf{z}_{i}\in\mathcal{B}\left(\widetilde{\boldsymbol{u}}_{k}\right)\right)-P_{\widetilde{\mathbf{z}}\left|\mathbf{z}'\right.}\left(\widetilde{\boldsymbol{u}}_{k}\left|\widetilde{\boldsymbol{v}}_{j}\right.\right)\right)\mathfrak{1}_{i'}\left(\widetilde{\ensuremath{\mathbf{z}'}}_{i'}=\widetilde{\boldsymbol{v}}_{j'}\right)\left(\mathfrak{1}_{i'}\left(\mathbf{z}_{i'}\in\mathcal{B}\left(\widetilde{\boldsymbol{u}}_{k}\right)\right)-P_{\widetilde{\mathbf{z}}\left|\mathbf{z}'\right.}\left(\widetilde{\boldsymbol{u}}_{k}\left|\widetilde{\boldsymbol{v}}_{j'}\right.\right)\right)\right)}{n^{2}}\nonumber \\
 & <\frac{1}{n}+\frac{\sum_{i=1}^{n}\sum_{i'=1,i\neq i}^{n}E\left(\mathfrak{1}_{i}\left(\widetilde{\ensuremath{\mathbf{z}'}}_{i}=\widetilde{\boldsymbol{v}}_{j}\right)\mathfrak{1}_{i'}\left(\widetilde{\ensuremath{\mathbf{z}'}}_{i'}=\widetilde{\boldsymbol{v}}_{j'}\right)\mathfrak{1}_{i}\left(\mathbf{z}_{i}\in\mathcal{B}\left(\widetilde{\boldsymbol{u}}_{k}\right)\right)\mathfrak{1}_{i'}\left(\mathbf{z}_{i'}\in\mathcal{B}\left(\widetilde{\boldsymbol{u}}_{k}\right)\right)\right)}{n^{2}}\nonumber \\
 & -\frac{\sum_{i=1}^{n}\sum_{i'=1,i\neq i}^{n}E\left(\mathfrak{1}_{i}\left(\widetilde{\ensuremath{\mathbf{z}'}}_{i}=\widetilde{\boldsymbol{v}}_{j}\right)\mathfrak{1}_{i'}\left(\widetilde{\ensuremath{\mathbf{z}'}}_{i'}=\widetilde{\boldsymbol{v}}_{j'}\right)\mathfrak{1}_{i}\left(\mathbf{z}_{i}\in\mathcal{B}\left(\widetilde{\boldsymbol{u}}_{k}\right)\right)P_{\widetilde{\mathbf{z}}\left|\mathbf{z}'\right.}\left(\widetilde{\boldsymbol{u}}_{k}\left|\widetilde{\boldsymbol{v}}_{j'}\right.\right)\right)}{n^{2}}\nonumber \\
 & -\frac{\sum_{i=1}^{n}\sum_{i'=1,i\neq i}^{n}E\left(\mathfrak{1}_{i}\left(\widetilde{\ensuremath{\mathbf{z}'}}_{i}=\widetilde{\boldsymbol{v}}_{j}\right)\mathfrak{1}_{i'}\left(\widetilde{\ensuremath{\mathbf{z}'}}_{i'}=\widetilde{\boldsymbol{v}}_{j'}\right)\mathfrak{1}_{i'}\left(\mathbf{z}_{i'}\in\mathcal{B}\left(\widetilde{\boldsymbol{u}}_{k}\right)\right)P_{\widetilde{\mathbf{z}}\left|\mathbf{z}'\right.}\left(\widetilde{\boldsymbol{u}}_{k}\left|\widetilde{\boldsymbol{v}}_{j}\right.\right)\right)}{n^{2}}\nonumber \\
 & +\frac{\sum_{i=1}^{n}\sum_{i'=1,i\neq i}^{n}E\left(\mathfrak{1}_{i}\left(\widetilde{\ensuremath{\mathbf{z}'}}_{i}=\widetilde{\boldsymbol{v}}_{j}\right)\mathfrak{1}_{i'}\left(\widetilde{\ensuremath{\mathbf{z}'}}_{i'}=\widetilde{\boldsymbol{v}}_{j'}\right)P_{\widetilde{\mathbf{z}}\left|\mathbf{z}'\right.}\left(\widetilde{\boldsymbol{u}}_{k}\left|\widetilde{\boldsymbol{v}}_{j'}\right.\right)P_{\widetilde{\mathbf{z}}\left|\mathbf{z}'\right.}\left(\widetilde{\boldsymbol{u}}_{k}\left|\widetilde{\boldsymbol{v}}_{j}\right.\right)\right)}{n^{2}}\nonumber \\
 & = \frac{1}{n}+\frac{\sum_{i=1}^{n}\sum_{i'=1,i\neq i}^{n}P_{\widetilde{\ensuremath{\mathbf{z}'}}_{i},\widetilde{\ensuremath{\mathbf{z}'}}_{i'}}\left(\widetilde{\boldsymbol{v}}_{j},\widetilde{\boldsymbol{v}}_{j'}\right)G_{i,i',j,j'}}{n^{2}}.
\end{align}
}}
$G_{i,i',j,j'}$ is given by
{\small{\begin{align}\label{G}
 & G_{i,i',j,j'}=P_{\mathbf{z}_{i},\mathbf{z}_{i'}\left|\widetilde{\ensuremath{\mathbf{z}'}}_{i},\widetilde{\ensuremath{\mathbf{z}'}}_{i'}\right.}\left(\mathbf{z}_{i}\in\mathcal{B}\left(\widetilde{\boldsymbol{u}}_{k}\right),\mathbf{z}_{i'}\in\mathcal{B}\left(\widetilde{\boldsymbol{u}}_{k}\right)\left|\widetilde{\boldsymbol{v}}_{j},\widetilde{\boldsymbol{v}}_{j'}\right.\right)-P_{\widetilde{\mathbf{z}}\left|\mathbf{z}'\right.}\left(\widetilde{\boldsymbol{u}}_{k}\left|\widetilde{\boldsymbol{v}}_{j'}\right.\right)P_{\mathbf{z}_{i}\left|\widetilde{\ensuremath{\mathbf{z}'}}_{i},\widetilde{\ensuremath{\mathbf{z}'}}_{i'}\right.}\left(\mathbf{z}_{i}\in\mathcal{B}\left(\widetilde{\boldsymbol{u}}_{k}\right)\left|\widetilde{\boldsymbol{v}}_{j},\widetilde{\boldsymbol{v}}_{j'}\right.\right)\nonumber \\
 & -P_{\widetilde{\mathbf{z}}\left|\mathbf{z}'\right.}\left(\widetilde{\boldsymbol{u}}_{k}\left|\widetilde{\boldsymbol{v}}_{j}\right.\right)P_{\mathbf{z}_{i'}\left|\widetilde{\ensuremath{\mathbf{z}'}}_{i},\widetilde{\ensuremath{\mathbf{z}'}}_{i'}\right.}\left(\mathbf{z}_{i'}\in\mathcal{B}\left(\widetilde{\boldsymbol{u}}_{k}\right)\left|\widetilde{\boldsymbol{v}}_{j},\widetilde{\boldsymbol{v}}_{j'}\right.\right)+P_{\widetilde{\mathbf{z}}\left|\mathbf{z}'\right.}\left(\widetilde{\boldsymbol{u}}_{k}\left|\widetilde{\boldsymbol{v}}_{j}\right.\right)P_{\widetilde{\mathbf{z}}\left|\mathbf{z}'\right.}\left(\widetilde{\boldsymbol{u}}_{k}\left|\widetilde{\boldsymbol{v}}_{j'}\right.\right).
\end{align}}}
For further bounding on $G_{i,i',j,j'}$, we note {{\small
\begin{align*}
& P_{\mathbf{z}_{i},\mathbf{z}_{i'}\left|\widetilde{\ensuremath{\mathbf{z}'}}_{i},\widetilde{\ensuremath{\mathbf{z}'}}_{i'}\right.}\left(\mathbf{z}_{i}\in\mathcal{B}\left(\widetilde{\boldsymbol{u}}_{k}\right),\mathbf{z}_{i'}\in\mathcal{B}\left(\widetilde{\boldsymbol{u}}_{k}\right)\left|\widetilde{\boldsymbol{v}}_{j},\widetilde{\boldsymbol{v}}_{j'}\right.\right)=\\&\frac{\int_{\boldsymbol{v}_{i}\in\mathcal{B}\left(\widetilde{\boldsymbol{v}}_{j}\right)}\int_{\boldsymbol{v}_{i'}\in\mathcal{B}\left(\widetilde{\boldsymbol{v}}_{j'}\right)}P_{\widetilde{\mathbf{z}}\left|\mathbf{z}'\right.}\left(\widetilde{\boldsymbol{u}}_{k}\left|\boldsymbol{v}_{i}\right.\right)P_{\widetilde{\mathbf{z}}\left|\mathbf{z}'\right.}\left(\widetilde{\boldsymbol{u}}_{k}\left|\boldsymbol{v}_{i'}\right.\right)f_{\mathbf{z}'_{i},\mathbf{z}'_{i'}}\left(\boldsymbol{v}_{i},\boldsymbol{v}_{i'}\right)d\boldsymbol{v}_{i}d\boldsymbol{v}_{i'}}{\int_{\boldsymbol{v}_{i}\in\mathcal{B}\left(\widetilde{\boldsymbol{v}}_{j}\right)}\int_{\boldsymbol{v}_{i'}\in\mathcal{B}\left(\widetilde{\boldsymbol{v}}_{j'}\right)}f_{\mathbf{z}'_{i},\mathbf{z}'_{i'}}\left(\boldsymbol{v}_{i},\boldsymbol{v}_{i'}\right)d\boldsymbol{v}_{i}d\boldsymbol{v}_{i'}}
\end{align*}
}} and {{\small
\begin{align}
 & P_{\mathbf{z}_{i}\left|\widetilde{\ensuremath{\mathbf{z}'}}_{i},\widetilde{\ensuremath{\mathbf{z}'}}_{i'}\right.}\left(\mathbf{z}_{i}\in\mathcal{B}\left(\widetilde{\boldsymbol{u}}_{k}\right)\left|\widetilde{\boldsymbol{v}}_{j},\widetilde{\boldsymbol{v}}_{j'}\right.\right)=  \frac{\int_{\boldsymbol{v}_{i}\in\mathcal{B}\left(\widetilde{\boldsymbol{v}}_{j}\right)}\int_{\boldsymbol{v}_{i'}\in\mathcal{B}\left(\widetilde{\boldsymbol{v}}_{j'}\right)}P_{\widetilde{\mathbf{z}}\left|\mathbf{z}'\right.}\left(\widetilde{\boldsymbol{u}}_{k}\left|\boldsymbol{v}_{i}\right.\right)f_{\mathbf{z}'_{i},\mathbf{z}'_{i'}}\left(\boldsymbol{v}_{i},\boldsymbol{v}_{i'}\right)d\boldsymbol{v}_{i}d\boldsymbol{v}_{i'}}{\int_{\boldsymbol{v}_{i}\in\mathcal{B}\left(\widetilde{\boldsymbol{v}}_{j}\right)}\int_{\boldsymbol{v}_{i'}\in\mathcal{B}\left(\widetilde{\boldsymbol{v}}_{j'}\right)}f_{\mathbf{z}'_{i},\mathbf{z}'_{i'}}\left(\boldsymbol{v}_{i},\boldsymbol{v}_{i'}\right)d\boldsymbol{v}_{i}d\boldsymbol{v}_{i'}}.
\end{align}
}}Hence, we have {{\small
\begin{align}
 & P_{\mathbf{z}_{i},\mathbf{z}_{i'}\left|\widetilde{\ensuremath{\mathbf{z}'}}_{i},\widetilde{\ensuremath{\mathbf{z}'}}_{i'}\right.}\left(\mathbf{z}_{i}\in\mathcal{B}\left(\widetilde{\boldsymbol{u}}_{k}\right),\mathbf{z}_{i'}\in\mathcal{B}\left(\widetilde{\boldsymbol{u}}_{k}\right)\left|\widetilde{\boldsymbol{v}}_{j},\widetilde{\boldsymbol{v}}_{j'}\right.\right)-P_{\widetilde{\mathbf{z}}\left|\mathbf{z}'\right.}\left(\widetilde{\boldsymbol{u}}_{k}\left|\widetilde{\boldsymbol{v}}_{j'}\right.\right)P_{\mathbf{z}_{i}\left|\widetilde{\ensuremath{\mathbf{z}'}}_{i},\widetilde{\ensuremath{\mathbf{z}'}}_{i'}\right.}\left(\mathbf{z}_{i}\in\mathcal{B}\left(\widetilde{\boldsymbol{u}}_{k}\right)\left|\widetilde{\boldsymbol{v}}_{j},\widetilde{\boldsymbol{v}}_{j'}\right.\right)\leq\nonumber \\
 & \max_{\boldsymbol{v}_{i}\in\mathcal{B}\left(\widetilde{\boldsymbol{v}}_{j}\right),\boldsymbol{v}_{i'}\in\mathcal{B}\left(\widetilde{\boldsymbol{v}}_{j'}\right)}P_{\widetilde{\mathbf{z}}\left|\mathbf{z}'\right.}\left(\widetilde{\boldsymbol{u}}_{k}\left|\boldsymbol{v}_{i}\right.\right)P_{\widetilde{\mathbf{z}}\left|\mathbf{z}'\right.}\left(\widetilde{\boldsymbol{u}}_{k}\left|\boldsymbol{v}_{i'}\right.\right)-\min_{\boldsymbol{v}_{i}\in\mathcal{B}\left(\widetilde{\boldsymbol{v}}_{j}\right),\boldsymbol{v}_{i'}\in\mathcal{B}\left(\widetilde{\boldsymbol{v}}_{j'}\right)}P_{\widetilde{\mathbf{z}}\left|\mathbf{z}'\right.}\left(\widetilde{\boldsymbol{u}}_{k}\left|\boldsymbol{v}_{i}\right.\right)P_{\widetilde{\mathbf{z}}\left|\mathbf{z}'\right.}\left(\widetilde{\boldsymbol{u}}_{k}\left|\boldsymbol{v}_{i'}\right.\right)
\end{align}
}}According to the Cauchy mean value theorem, for closed domain $\mathcal{B}\left(\widetilde{\boldsymbol{v}}_{j}\right)$
and $\mathcal{B}\left(\widetilde{\boldsymbol{v}}_{j'}\right)$, we
have {{\small
\begin{equation}
 \max_{\boldsymbol{v}_{i}\in\mathcal{B}\left(\widetilde{\boldsymbol{v}}_{j}\right),\boldsymbol{v}_{i'}\in\mathcal{B}\left(\widetilde{\boldsymbol{v}}_{j'}\right)}P_{\widetilde{\mathbf{z}}\left|\mathbf{z}'\right.}\left(\widetilde{\boldsymbol{u}}_{k}\left|\boldsymbol{v}_{i}\right.\right)P_{\widetilde{\mathbf{z}}\left|\mathbf{z}'\right.}\left(\widetilde{\boldsymbol{u}}_{k}\left|\boldsymbol{v}_{i'}\right.\right)- \min_{\boldsymbol{v}_{i}\in\mathcal{B}\left(\widetilde{\boldsymbol{v}}_{j}\right),\boldsymbol{v}_{i'}\in\mathcal{B}\left(\widetilde{\boldsymbol{v}}_{j'}\right)}P_{\widetilde{\mathbf{z}}\left|\mathbf{z}'\right.}\left(\widetilde{\boldsymbol{u}}_{k}\left|\boldsymbol{v}_{i}\right.\right)P_{\widetilde{\mathbf{z}}\left|\mathbf{z}'\right.}\left(\widetilde{\boldsymbol{u}}_{k}\left|\boldsymbol{v}_{i'}\right.\right)
 \leq c_{1}\delta_2^{N_L+N_M}
\end{equation}
}}where $c_{1}$ is a constant that depends on $P_{\widetilde{\mathbf{z}}\left|\mathbf{z}'\right.}\left(\cdot\left|\cdot\right.\right)$.
Similarly, we can obtain {{\small
\begin{align*}
 P_{\widetilde{\mathbf{z}}\left|\mathbf{z}'\right.}\left(\widetilde{\boldsymbol{u}}_{k}\left|\widetilde{\boldsymbol{v}}_{j}\right.\right)P_{\widetilde{\mathbf{z}}\left|\mathbf{z}'\right.}\left(\widetilde{\boldsymbol{u}}_{k}\left|\widetilde{\boldsymbol{v}}_{j'}\right.\right)-P_{\widetilde{\mathbf{z}}\left|\mathbf{z}'\right.}\left(\widetilde{\boldsymbol{u}}_{k}\left|\widetilde{\boldsymbol{v}}_{j}\right.\right)P_{\mathbf{z}_{i'}\left|\widetilde{\ensuremath{\mathbf{z}'}}_{i},\widetilde{\ensuremath{\mathbf{z}'}}_{i'}\right.}\left(\mathbf{z}_{i'}\in\mathcal{B}\left(\widetilde{\boldsymbol{u}}_{k}\right)\left|\widetilde{\boldsymbol{v}}_{j},\widetilde{\boldsymbol{v}}_{j'}\right.\right) \leq c_{2}\delta_2^{N_L+N_M}
\end{align*}
}}for closed domain $\mathcal{B}\left(\widetilde{\boldsymbol{v}}_{j}\right)$
and $\mathcal{B}\left(\widetilde{\boldsymbol{v}}_{j'}\right)$, $c_{2}$
is a constant depends on $P_{\widetilde{\mathbf{z}}\left|\mathbf{z}'\right.}\left(\cdot\left|\cdot\right.\right)$.
Therefore, $G_{i,i',j,j'}$ can be bound as
\begin{equation}
G_{i,i',j,j'}\leq\left(c_{1}+c_{2}\right)\delta_2^{N_L+N_M}.
\end{equation}
We define a set $\mathcal{J}$, where $\left\{ j,j'\right\} \in\mathcal{J}$
means $\mathcal{B}\left(\widetilde{\boldsymbol{v}}_{j}\right)$ and
$\mathcal{B}\left(\widetilde{\boldsymbol{v}}_{j'}\right)$ are closed
domains. On the other hand, notice that the alphabets are finite, $\Re\left\{ \mathcal{Z}\right\} \subseteq\left[-\alpha_{1},\alpha_{1}-\delta_2\right]^{N_{M}+N_{L}}$,
$\Im\left\{ \mathcal{Z}\right\} \subseteq\left[-\alpha_{1},\alpha_{1}-\delta_2\right]^{N_{M}+N_{L}}$, and thus $\left\{ j,j'\right\} \in\overline{\mathcal{J}}$, $P_{\widetilde{\ensuremath{\mathbf{z}'}}_{i},\widetilde{\ensuremath{\mathbf{z}'}}_{i'}}\left(\widetilde{\boldsymbol{v}}_{j},\widetilde{\boldsymbol{v}}_{j'}\right)=0$ as the quantization ranges approach infinity.
Therefore, (\ref{fupper4}) becomes
{{\tiny
\begin{align}\label{fupper7}
 & \Pr\left\{ \left|\sum_{j=1}^{\left|\widetilde{\mathcal{Z}}'\right|}{\underbrace{\left(\frac{\sum_{i=1}^{n}\mathfrak{1}_{i}\left(\widetilde{\ensuremath{\mathbf{z}'}}_{i}=\widetilde{\boldsymbol{v}}_{j}\right)\mathfrak{1}_{i}\left(\mathbf{z}_{i}\in\mathcal{B}\left(\widetilde{\boldsymbol{u}}_{k}\right)\right)}{n}-\frac{\sum_{i=1}^{n}\mathfrak{1}_{i}\left(\widetilde{\ensuremath{\mathbf{z}'}}_{i}=\widetilde{\boldsymbol{v}}_{j}\right)}{n}P_{\widetilde{\mathbf{z}}\left|\mathbf{z}'\right.}\left(\widetilde{\boldsymbol{u}}_{k}\left|\widetilde{\boldsymbol{v}}_{j}\right.\right)\right)}_{H_{j}}}\right|>\mu\right\}\leq\frac{1}{\mu^{2}}\sum_{j=1}^{{\left|\widetilde{\mathcal{Z}}'\right|}}\sum_{j'=1}^{{\left|\widetilde{\mathcal{Z}}'\right|}}E\left(H_{j}H_{j'}\right)\\\nonumber
 & =\frac{1}{\mu^{2}}\left(\frac{\left|\widetilde{\mathcal{Z}}'\right|^{2}}{n}+\sum_{\left\{ j,j'\right\} \in\mathcal{J}}^ {}\frac{\sum_{i=1}^{n}\sum_{i\text{'}=1,i\neq i}^{n}P_{\widetilde{\ensuremath{\mathbf{z}'}}_{i},\widetilde{\ensuremath{\mathbf{z}'}}_{i'}}\left(\widetilde{\boldsymbol{v}}_{j},\widetilde{\boldsymbol{v}}_{j'}\right)G_{i,i',j,j'}}{n^{2}}+\sum_{\left\{ j,j'\right\} \in\overline{\mathcal{J}}}^ {}\frac{\sum_{i=1}^{n}\sum_{i\text{'}=1,i\neq i}^{n}P_{\widetilde{\ensuremath{\mathbf{z}'}}_{i},\widetilde{\ensuremath{\mathbf{z}'}}_{i'}}\left(\widetilde{\boldsymbol{v}}_{j},\widetilde{\boldsymbol{v}}_{j'}\right)G_{i,i',j,j'}}{n^{2}}\right)\\\nonumber
 &
   \leq\frac{1}{\mu^{2}}\left(\frac{\left|\widetilde{\mathcal{Z}}'\right|^{2}}{n}+\sum_{\left\{ j,j'\right\} \in\mathcal{J}}^ {}\frac{n^{2}P_{\widetilde{\ensuremath{\mathbf{z}'}}_{i},\widetilde{\ensuremath{\mathbf{z}'}}_{i'}}\left(\widetilde{\boldsymbol{v}}_{j},\widetilde{\boldsymbol{v}}_{j'}\right)c\left(N_L+N_M\right)\delta_2^{N_L+N_M}}{n^{2}}+\frac{\sum_{i=1}^{n}\sum_{i\text{'}=1,i\neq i}^{n}\sum_{\left\{ j,j'\right\} \in\overline{\mathcal{J}}}^ {}P_{\widetilde{\ensuremath{\mathbf{z}'}}_{i},\widetilde{\ensuremath{\mathbf{z}'}}_{i'}}\left(\widetilde{\boldsymbol{v}}_{j},\widetilde{\boldsymbol{v}}_{j'}\right)}{n^{2}}\right)\\\nonumber
 & \leq\frac{1}{\mu^{2}}\left(\frac{\left|\widetilde{\mathcal{Z}}'\right|^{2}}{n}+\left(c_1+c_2\right)\delta_2^{N_L+N_M}+\frac{\sum_{i=1}^{n}\sum_{i\text{'}=1,i\neq i}^{n}\sum_{\left\{ j,j'\right\} \in\overline{\mathcal{J}}}^ {}P_{\widetilde{\ensuremath{\mathbf{z}'}}_{i},\widetilde{\ensuremath{\mathbf{z}'}}_{i'}}\left(\widetilde{\boldsymbol{v}}_{j},\widetilde{\boldsymbol{v}}_{j'}\right)}{n^{2}}\right)=\frac{1}{\mu^{2}}\left(\frac{\left|\widetilde{\mathcal{Z}}'\right|^{2}}{n}+\left(c_1+c_2\right)\delta_2^{N_L+N_M}\right)
\end{align}
}}and 
the proof is completed.
\end{IEEEproof}

\bibliographystyle{IEEEtran}

\begin{thebibliography}{10}
\bibitem{Prasad}
K. N. R. S. V. Prasad, E. Hossain and V. K. Bhargava, ``Energy Efficiency in Massive MIMO-Based 5G Networks: Opportunities and Challenges," \emph{\fontsize{10}{1.5} {IEEE Wireless Communications}}, vol.24, no.3, pp.86-94, 2017.





\bibitem{Shafi}
M. Shafi et al., ``5G: A Tutorial Overview of Standards, Trials, Challenges, Deployment, and Practice,"  \emph{IEEE Journal on Selected Areas in Communications}, vol. 35, no. 6, pp. 1201-1221, June 2017.

\bibitem{Liusurvey}
Y. Liu, H. H. Chen and L. Wang, ``Physical Layer Security for Next Generation Wireless Networks: Theories, Technologies, and Challenges,"  \emph{IEEE Communications Surveys \& Tutorials}, vol. 19, no. 1, pp. 347-376, Firstquarter 2017.


\bibitem{chensurvey}
X. Chen, D. W. K. Ng, W. H. Gerstacker and H. H. Chen, ``A Survey on Multiple-Antenna Techniques for Physical Layer Security," \emph{IEEE Communications Surveys \& Tutorials}, vol. 19, no. 2, pp. 1027-1053, Secondquarter 2017.

\bibitem{Hesurvey}
B. He, X. Zhou, T. D. Abhayapala, ``Wireless physical layer security with imperfect channel state information: A survey", \emph{ZTE Communications}, vol. 11, no. 3, pp. 11-19, Sep. 2013.

\bibitem{PLSMaMIMO}
D. Kapetanovic, G. Zheng and F. Rusek,  ``Physical layer security for massive MIMO: An overview on passive eavesdropping and active attacks," \emph{IEEE Communications Magazine}, vol. 53, no. 6, pp. 21-27, June 2015.


\bibitem{Xiong2015}
 Q. Xiong, Y. C. Liang, K. H. Li and Y. Gong, ``An Energy-Ratio-Based Approach for Detecting Pilot Spoofing Attack in Multiple-Antenna Systems," \emph{IEEE Transactions on Information Forensics and Security}, vol. 10, no. 5, pp. 932-940, May 2015.

 \bibitem{Kapetanovic2014}
D. Kapetanovic, A. Al-Nahari, A. Stojanovic and F. Rusek, ``Detection of active eavesdroppers in massive MIMO," in \emph{Proc. IEEE 25th Annual International Symposium on Personal, Indoor, and Mobile Radio Communication (PIMRC)}, Washington DC, June 2014, pp. 585-589.

\bibitem{Kang}
J. M. Kang, C. In and H. M. Kim, ``Detection of Pilot Contamination Attack for Multi-Antenna Based Secrecy Systems," in \emph{Proc. IEEE 81st Vehicular Technology Conference (VTC Spring)}, Glasgow, 2015, pp. 1-5.

\bibitem{CFOKnightly}
X. Zhang and E. W. Knightly, ``Massive MIMO pilot distortion attack and zero-startup-cost detection: Analysis and experiments," in \emph{Proc. 2017 IEEE Conference on Communications and Network Security (CNS)}, Las Vegas, NV, USA, 2017, pp. 1-9.

\bibitem{CFOZhang}
W. Zhang, H. Lin and R. Zhang, ``Detection of Pilot Contamination Attack based on Uncoordinated Frequency Shifts,"  \emph{IEEE Transactions on Communications}, Early Access, 2018.

\bibitem{Tugnait2016}
J. K. Tugnait, ``Detection of pilot contamination attack in T.D.D./S.D.M.A. systems," in \emph{Proc. IEEE International Conference on Acoustics, Speech and Signal Processing (ICASSP)}, Shanghai, May 2016, pp. 3576-3580.


\bibitem{TugnaitWL}
J. K. Tugnait, ``Self-Contamination for Detection of Pilot Contamination Attack in Multiple Antenna Systems,"  \emph{IEEE Wireless Communications Letters}, vol. 4, no. 5, pp. 525-528, Oct. 2015.

\bibitem{kkwong}
D. Kapetanovi, G. Zheng, K. K. Wong and B. Ottersten, ``Detection of pilot contamination attack using random training and massive MIMO," in \emph{Proc. IEEE 24th Annual International Symposium on Personal, Indoor, and Mobile Radio Communications (PIMRC)}, London, 2013, pp. 13-18.

\bibitem{VTCrandom}
X. Wang, M. Liu, D. Wang and C. Zhong, ``Pilot Contamination Attack Detection Using Random Symbols for Massive MIMO Systems," in \emph{Proc. 2017 IEEE 85th Vehicular Technology Conference (VTC Spring)}, Sydney, NSW, 2017, pp. 1-7.

\bibitem{WCSPrandom}
X. Hou, C. Gao, Y. Zhu and S. Yang, ``Detection of active attacks based on random orthogonal pilots," in \emph{Proc. 2016 8th International Conference on Wireless Communications \&\ Signal Processing (WCSP)}, Yangzhou, 2016, pp. 1-4.


\bibitem{Vinogradova}
J. Vinogradova, E. Bj\"{o}rnson and E. G. Larsson, ``Detection and mitigation of jamming attacks in massive MIMO systems using random matrix theory," in \emph{Proc. IEEE 17th International Workshop on Signal Processing Advances in Wireless Communications (SPAWC)}, Edinburgh, 2016, pp. 1-5.



\bibitem{ICCrandom}
J. Xie, Y. C. Liang, J. Fang and X. Kang, ``Two-stage uplink training for pilot spoofing attack detection and secure transmission," 2017 IEEE International Conference on Communications (ICC), Paris, 2017, pp. 1-6.

%


\bibitem{ICTC}
J. Park, S. Yun and J. Ha, ``Detection of pilot contamination attack in the MU-MISOME broadcast channels," in \emph{Proc. International Conference on Information and Communication Technology Convergence (ICTC)}, Jeju, 2016, pp. 664-666.






\bibitem{DCETcom}
J. Yang, S. Xie, X. Zhou, R. Yu and Y. Zhang, ``A Semiblind Two-Way Training Method for Discriminatory Channel Estimation in MIMO Systems,"  \emph{IEEE Transactions on Communications}, vol. 62, no. 7, pp. 2400-2410, July 2014.


\bibitem{Xiong2016}
Q. Xiong, Y. C. Liang, K. H. Li, Y. Gong and S. Han, ``Secure Transmission Against Pilot Spoofing Attack: A Two-Way Training-Based Scheme," \emph{IEEE Transactions on Information Forensics and Security}, vol. 11, no. 5, pp. 1017-1026, May 2016.




\bibitem{Bai}
F. Bai, P. Ren, Q. Du and L. Sun, ``A hybrid channel estimation strategy against pilot spoofing attack in MISO system," in \emph{Proc. IEEE 27th Annual International Symposium on Personal, Indoor, and Mobile Radio Communications (PIMRC)}, Valencia, 2016, pp. 1-6.

\bibitem{RenICC}
D. Xu, P. Ren, Y. Wang, Q. Du and L. Sun, ``ICA-SBDC: A channel estimation and identification mechanism for MISO-OFDM systems under pilot spoofing attack," in \emph{2017 IEEE International Conference on Communications (ICC)}, Paris, 2017, pp. 1-6.

\bibitem{TugnaitTcom}
J. K. Tugnait, ``On Detection and Mitigation of Reused Pilots in Massive MIMO Systems," \emph{IEEE Transactions on Communications}, vol. 66, no. 2, pp. 688-699, Feb. 2018.

\bibitem{TugnaitTcomattack}
J. K. Tugnait, ``Pilot Spoofing Attack Detection and Countermeasure,"  \emph{IEEE Transactions on Communications}, vol. 66, no. 5, pp. 2093-2106, May 2018.

\bibitem{TTDoPJA}
T. T. Do, E. Bj\"{o}rnson, E. G. Larsson and S. M. Razavizadeh,  ``Jamming resistant
receivers for the massive MIMO uplink," \emph{IEEE Transactions on Information Forensics and Security} vol. 13, no. 1, pp. 210-223, Jan. 2018.

\bibitem{CodeOFDM}
D. Xu, P. Ren, J. A. Ritcey and Y. Wang, ``Code-Frequency Block Group Coding for Anti-Spoofing Pilot Authentication in Multi-Antenna OFDM Systems," \emph{IEEE Transactions on Information Forensics and Security,} vol. 13, no. 7, pp. 1778-1793, July 2018.
\bibitem{WangPSJA}
H. M. Wang, K. W. Huang and T. A. Tsiftsis, ``Multiple Antennas Secure Transmission under Pilot Spoofing and Jamming Attack," \emph{IEEE Journal on Selected Areas in Communications}, vol. 36, no. 4, pp. 860-876, April 2018.
\bibitem{Trappe}
W. Trappe,``The challenges facing physical layer security," \emph{IEEE Communications Magazine}, vol. 53, no. 6, pp. 16-20, June 2015.

\bibitem{Confversion}
R. Cao, T. Wong, H. Gao, D. Wang, and Y. Lu, ``Blind Channel Direction Separation Against Pilot Spoofing Attack in Massive MIMO System", in \emph{2018 26th European 
Signal Processing Conference (EUSIPCO),} Roman, 2018, pp.2559-2563.
\bibitem{Muller2014}
R. R. M\"{u}ller, L. Cottatellucci and M. Vehkaper\"{a}, ``Blind Pilot Decontamination,"  \emph{IEEE Journal of Selected Topics in Signal Processing}, vol. 8, no. 5, pp. 773-786, Oct. 2014.
\bibitem{FastICAJ}
A. Hyvarinen, ``Fast and robust fixed-point methods for independent component analysis,'' \emph{IEEE Transactions on Neural Networks}, vol. 10, no. 3, pp. 626-634, May 1999.
\bibitem{FastICAC}
V. Zarzoso, P. Comon and M. Kallel, ``How fast is FastICA?'' \emph{2006 14th European Signal Processing Conference}, Florence, 2006, pp. 1-5.
\end{thebibliography}

\end{document}